\documentclass[8pt,twocolumn]{autart}    % Enable this
\usepackage{graphicx}

\newtheorem{assumption}{Assumption}
\usepackage{bm}
\usepackage{amsmath}
\usepackage{amssymb}
\usepackage{xcolor}
\graphicspath{{Figures/}}
\makeatletter
\let\c@author\relax
\makeatother
\usepackage[style=authoryear, maxnames=6, backend=biber]{biblatex}
\DefineBibliographyExtras{english}{%
}
\DeclareFieldFormat[article,book,incollection,inproceedings]{title}{#1}
\DeclareFieldFormat{volume}{#1} % No 'vol.' anywhere
\DeclareFieldFormat[article]{number}{\mkbibparens{#1}} % Parentheses around number
\DeclareFieldFormat[article]{pages}{#1} % Remove 'pp.'
\renewbibmacro{in:}{} % Remove 'in:' before journal or booktitle
\renewbibmacro*{in:}{
  \ifentrytype{inproceedings}
    {\printtext{\bibstring{in}\space}} % print 'in' only for inproceedings
    {} % print nothing for all others
}

\renewbibmacro*{volume+number+eid}{
  \setunit{\addcomma\space}% Comma before volume
  \printfield{volume}%
  \printfield{number}% No space between volume and number
  \setunit{\addcomma\space}%
  \printfield{eid}
}

\usepackage{xcolor}
\usepackage{marginnote}

\definecolor{magenta}{rgb}{0.8, 0.2, 0.46}
\definecolor{orange}{rgb}{1.5, 0.5, 0}
\definecolor{turquesa}{rgb}{0, 0.7, 1}

\def\BibTeX{{\rm B\kern-.05em{\sc i\kern-.025em b}\kern-.08em
		T\kern-.1667em\lower.7ex\hbox{E}\kern-.125emX}}

\newcommand{\bfx}{\mbox{$\bm{x}$}}

\def\lef[{\left[\begin{array}}
\def\rig]{\end{array}\right]}
\def\qed{\hfill$\Box \Box \Box$}
\def\bfx{{\bf x}}

\def\rea{\mathbb{R}}

\usepackage{psfrag}

\def\rea{\mathbb{R}}
\def\begequ{\begin{equation}}
\def\endequ{\end{equation}}
\def\lab{\label}
\def\begite{\begin{itemize}}
\def\endite{\end{itemize}}
\def\begarr{\begin{array}}
\def\endarr{\end{array}}
\def\begequarr{\begin{eqnarray}}
\def\endequarr{\end{eqnarray}}

\usepackage{multicol}

\def\liminf{\lim_{t \to \infty}}

\def\liminf{\lim_{t \to \infty}}

\def\L2{{\cal L}_2}
\def\L2e{{\cal L}_{2e}}

\def\rea{\mathbb{R}}

\def\sign{\mbox{sign}}

\def\sign{\mbox{sign}}

\def\et{\varepsilon_t}

\def\begmat#1{\begin{bmatrix}#1\end{bmatrix}}
\def\begali#1{\begin{align}{#1}\end{align}}
\def\begalis#1{\begin{align*}{#1}\end{align*}}
\def\begsubequ{\begin{subequations}}
\def\endsubequ{\end{subequations}}
\def\begequarr{\begin{eqnarray}}
\def\endequarr{\end{eqnarray}}
\def\begequarrs{\begin{eqnarray*}}
\def\endequarrs{\end{eqnarray*}}
\def\begarr{\begin{array}}
\def\endarr{\end{array}}
\def\begequ{\begin{equation}}
\def\endequ{\end{equation}}
\def\lab{\label}
\def\begdes{\begin{description}}
\def\enddes{\end{description}}
\def\begenu{\begin{enumerate}}
\def\begite{\begin{itemize}}
\def\endite{\end{itemize}}
\def\endenu{\end{enumerate}}

\def\lef[{\left[\begin{array}}
\def\rig]{\end{array}\right]}
\def\qed{\hfill$\Box \Box \Box$}
\def\begcen{\begin{center}}
\def\endcen{\end{center}}
\def\begrem{\begin{remark}\rm}
\def\endrem{\end{remark}}
\def\begassums{\begin{assumption*}}
\def\endassums{\end{assumption*}}
\def\begassu{\begin{ass}}
\def\endassu{\end{ass}}
\def\beglem{\begin{lemma}}
\def\endlem{\end{lemma}}
\def\begcor{\begin{corollary}}
\def\endcor{\end{corollary}}
\def\begfac{\begin{fact}}
\def\endfac{\end{fact}}
\def\begass{\begin{assumption}}
\def\endass{\end{assumption}}

\def\begmat#1{\begin{bmatrix}#1\end{bmatrix}}
\def\begali#1{\begin{align}{#1}\end{align}}
\def\begalis#1{\begin{align*}{#1}\end{align*}}
\usepackage{color}

\begin{document}
\begin{frontmatter}
\title{An Experimental Comparison of Sliding Mode and Immersion and Invariance Adaptive Controllers for Position-feedback Tracking of a Simple Mechanical System with Friction}
%\thanks[footnoteinfo]{Corresponding author L. Cervantes-Pérez (d.lecervantesp@lalaguna.tecnm.mx).}
\author[1]{Luis Cervantes-Pérez}\ead{d.lecervantesp@lalaguna.tecnm.mx},
\author[1]{Víctor Santibáñez}\ead{vasantibanezd@lalaguna.tecnm.mx},
\author[2]{Jesús Sandoval}\ead{jesus.sg@lapaz.tecnm.mx},
\author[3]{Romeo Ortega}\ead{romeo.ortega@itam.mx},
\author[3]{Jose Guadalupe Romero}\ead{jose.romerovelazquez@itam.mx},
\address[1]{Tecnológico Nacional de México / I.T. de La Laguna, Blvd. Revolución y, Av. Instituto Tecnológico de La Laguna s/n, Torreón, Coahuila, México, 27000.}  % Please supply
\address[2]{Tecnológico Nacional de México / I.T. de La Paz, Blvd. Forjadores de Baja California Sur 4720, La Paz, Baja California Sur, México, 23080.}
\address[3]{Departamento de Ingeniería Eléctrica y Electrónica / I.T. Autónomo de México, Río Hondo 1, CDMX, México, 01080.}

\begin{abstract}
The purpose of this paper is to illustrate, in an experimental facility consisting of a simple pendular device, the performance of  a sliding mode adaptive position-feedback tracking controller of mechanical systems with friction reported in the literature. To put this experimental evidence in perspective, we compare the performance of the sliding mode scheme with the one obtained by an adaptive controller designed following the well-known immersion and invariance technique.
\end{abstract}
\end{frontmatter}

\section{Introduction}
 \label{sec1}
%%%%%%%%%%%%
%
It is widely accepted in the control community that it is undesirable to differentiate signals and to inject high-gain (HG) in a control loop. It is well-known that both actions have the deleterious effects of reduction of the stability margins and noise amplification, the latter being  unavoidable in any practical application, in particular, in mechanical systems. Both operations are intrinsic to sliding mode (SM) designs, that have proliferated in the theoretical control literature in the last few years.
\par A critical discussion on the ``{\em sui géneris}" control theoretic scenario adopted for the application of some SM designs is carried out in \cite{ReflectionsSM}. Namely, the system dynamics is rewritten as the cascade connection of a very simple part---usually a linear time invariant plant with known parameters---which is ``perturbed" by an  additive {\em ``disturbance"} that is assumed to satisfy some Lipschitz-like condition.  The latter assumption allows the designer to propose a HG mechanism, {\em e.g.}, an infinite gain relay operator, that {\em ``dominates"} the ``disturbance". To accomplish the latter task, it is assumed that the signals are {\em bounded}, invoking the specious assumption that the system lives in some compact set. It is argued in  \cite{ReflectionsSM} that, neither the assumed scenario nor the standing assumptions, are compatible with current engineering practice in a large variety of applications. Experimental evidence of  the aforementioned deleterious effects of HG injection were reported over forty years ago in \cite{ORT}, and more recently in \cite{ARAetal}.
\par Our objective in this paper is to bring again to the readers attention these facts. This time we illustrate it with an experimental facility consisting of a simple pendular device, for which we implement a SM adaptive position-feedback tracking controller of mechanical systems with friction reported in  \cite{SM1}. To put this experimental evidence in perspective, we compare the performance of the SM scheme with the one obtained by the adaptive controller reported in \cite{rom2025adaptive}, which is designed following the well-known immersion and invariance (I\&I) technique \cite{astolfi2008nonlinear}.

\par The remainder of our paper is organized as follows. In Section \ref{sec2} we describe the experimental facility and explain the tasks that we wish to accomplish. In Section \ref{sec3} we present the speed observers and adaptive controllers to be tested experimentally. Sections \ref{sec4} and \ref{sec5} are devoted to the presentation of the experimental results, for the speed observation and trajectory tracking tasks, respectively. A final set of comparative simulation results on a hydro-mechanical system reported in \cite{SM2} for the SM design and in \cite{rom2025adaptive} for an I$\&$I one is presented in Section \ref{sec6}. We wrap-up the paper with some concluding remarks in Section \ref{sec7}.
%
%%%%%%%%%%%%%%%%%
\section{Description of the Experimental Facility and Tests}
\lab{sec2}
%%%%%%%%%%%%%%%
%
\subsection{Experimental set up}
\begin{figure}[h!]
  \centering
  \includegraphics[width=0.4\textwidth]{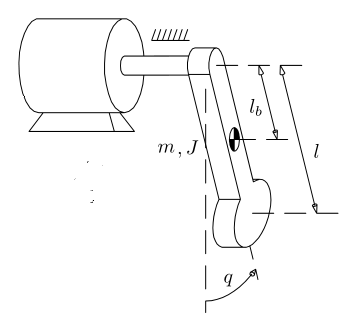}
  \caption{Sketch of the pendular device.}
    \label{fig1}
\end{figure}

\lab{subsec21}
%%%%%%%%%%%%%%%
%
We consider in this paper the pendular device depicted in Fig. \ref{fig1}, consisting of a direct-drive  arm with a single vertical link.

We assume that the device is subject to a friction force, whose behavior is approximated by Coulomb and static friction \cite{ARMetal}.

\par This leads to the following well-known dynamic model of the pendular device
\begin{equation}\label{sys}
J \ddot q + {\theta_1} \dot q+{\theta_2}\tanh(\vartheta \dot q)+m l_b g \sin(q)= u,
\end{equation}
where $q(t) \in [0,2\pi)$ is the pendulum angle, $u(t) \in \rea$ is the input torque, $\theta_1>0$ is the viscous friction coefficient, $\theta_2>0$ is a Coulomb friction coefficient, whose relay model is approximated with a $\tanh(\cdot)$ function---with $\vartheta>0$ a large number---and the positive constants $J,m,g$ and $l_b$ are physical parameters, as depicted in Fig. \ref{fig1}, which are assumed to be {\em known}. To simplify the notation we define $x_1:=q$ and $x_2:=\dot q$.

\begin{figure}[h!]
  \centering
  \includegraphics[scale=0.25]{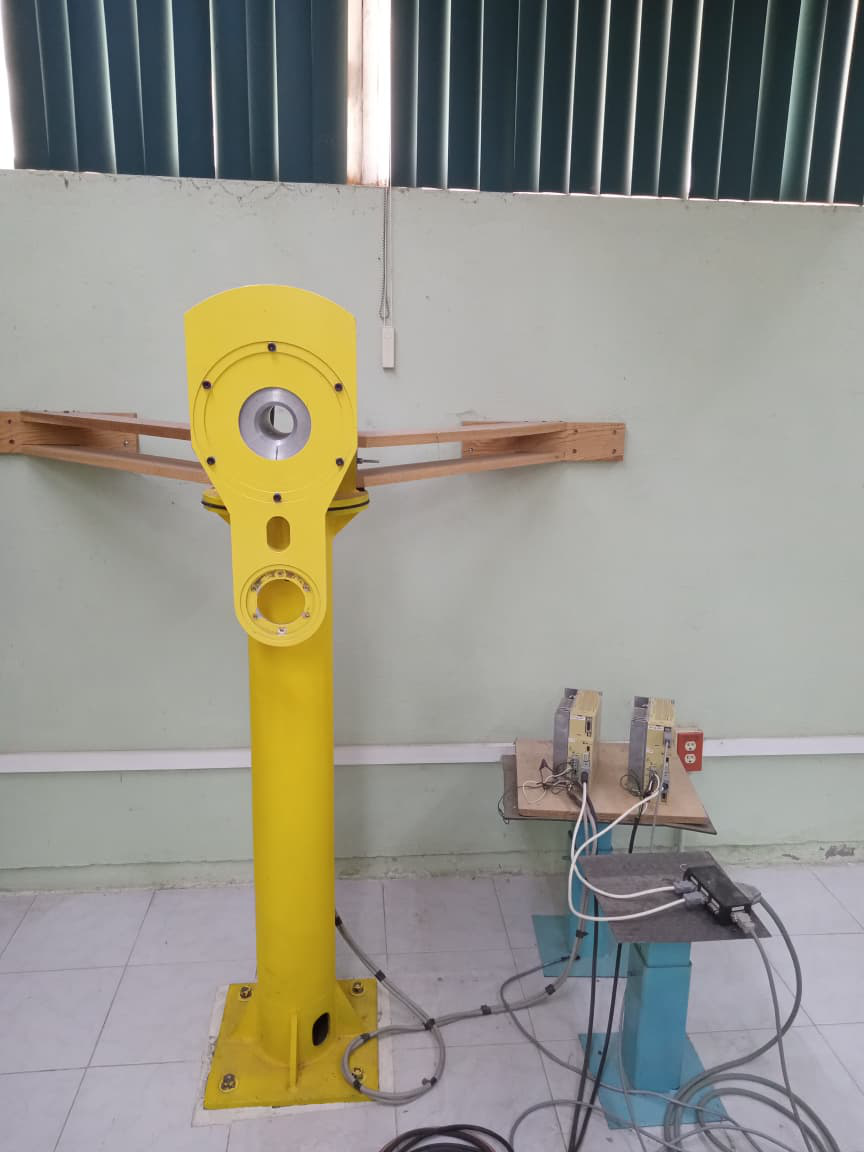}
  \caption{Pendular device.}
    \label{fig2}
\end{figure}

The experimental device was designed and built at the La Laguna Technological Institute, Mexico. A picture of the pendular device is shown in Fig.~\ref{fig2}.

The joint is actuated by a direct-drive DC motor manufactured by Yokogawa Electric Corporation, model \textit{DM1A200G}. The motor operates in torque mode, providing a maximum torque of $\pm200~\mathrm{[Nm]}$. It is also equipped with a high precision relative encoder offering a resolution of $1{,}024{,}000$ pulses per revolution. The proposed controller was implemented digitally in MATLAB using the Simulink Desktop Real-Time environment, with a sampling period of $1~\mathrm{[ms]}$. The experiments were conducted on a PC running Windows~10, equipped with an \textit{Intel Core i5-14600K} CPU, $32~\mathrm{GB}$ of RAM, an RTX~3060 GPU, and a $512~\mathrm{GB}$ M.2 SSD. Communication between the PC and the robot was established through a Sensoray~626 data acquisition card (DAQ) installed in the PCI port of the motherboard. The DAQ interfaced with the motor driver (PARKER COMPUMOTOR model DMG3-1200A) by sending analog voltage signals corresponding to the control torques. In turn, the driver transmitted pulse trains replicating the motor encoder signals to the DAQ digital inputs.

The physical parameters of the link are listed in Table~\ref{tab:parrob}, and they were originally reported in \cite{Gamez,Gandarilla}. The table summarizes the meaning of the parameters and their numerical values. The number $\vartheta$ was changed in the various experiments, and is given in the corresponding sections.

\begin{table*}
	\caption{Physical parameters of the pendular device.}
	\label{tab:parrob}
	\begin{center}
		\begin{tabular}{cccc} \hline
			Description  & Notation & Value & Units \\ \hline
			Length of Link  & $l$ & 0.313 & $\mathrm{m}$ \\
			Distance to the center of mass  & $l_{b}$ & 0.0641 & $\mathrm{m}$  \\
		Mass of Link  & $m$ & 22.4466 & $\mathrm{K_g}$ \\
		   Inertia relative to center of mass  & $J$ & 0.7013  & $\rm Kg\ m^2$ \\
          Viscous friction coefficient & $\theta_1$ & $5.317$ &  $\mathrm{Nm \ s/rad}$ \\
          Coulomb friction coefficient & $\theta_2$ & $11.6403$ &  $\mathrm{Nm }$ \\
			Gravity acceleration constant & $g$ & 9.81 & $\frac{{\rm m}}{{\rm s^2}}$ \\ \hline
		\end{tabular}
	\end{center}
\end{table*}
\subsection{Experimental tests}
\lab{subsec22}
%%%%%%%%%%%%%%%
%
We tried experimentally two speed observers  and associated adaptive tracking controllers: \\

\noindent {\bf (i)} the one based on I\&I recently reported in \cite{rom2025adaptive};\\

\noindent {\bf (ii)} the SM scheme described in  \cite{SM1}.\\

For these two schemes we carried-out the following experimental tests.
\begenu
\item[{\bf T1}] Open-loop test of the observers for two different input signals:
$$
u(t)=
\begin{cases}
25 \sin(5t) \\
14 \;\sign\big(\sin({\pi t\over 3})\big).
\end{cases}
$$
\item[{\bf T2}] Adaptive tracking test for the position reference signal:
$$
x_{1d}(t)=0.3\big[1-e^{-2.0t^3}\sin(7t) \big].
$$
\endenu

The purpose of trying different inputs and angle reference signals is to see the behavior of the schemes in the face of different {\em excitation} conditions. Also, being aware of the impact the choice of the tuning gains has on the performance of the algorithms, we carry out a discussion on their tuning and show the behavior for various values of these parameters. Unfortunately, as thoroughly discussed in \cite{ReflectionsSM}, guidelines for the tuning of SM algorithms are almost always conspicuous by their absence---usually simply saying that the gains have to be taken {\em ``sufficiently" large}.
%
%%%%%%%%%%%%%%%%%
\section{Proposed Adaptive Speed Observers and Controllers}
\lab{sec3}
%%%%%%%%%%%%%%%5
%
In this section we present the mathematical description of the schemes that we will try experimentally as described above. We assume that only the angle $q$ is measurable and propose to {\em cancel} the known gravity force, both, in the observers and in the tracking controllers.

Given the {\em estimated} state vector $\hat \bfx(t) \in \rea^2$, we denote the state {\em observation error} as $\tilde \bfx:=\hat \bfx -\bfx$. On the other hand, for a given position {\em reference} signal $x_{1d}(t) \in \rea$, we define the position error $e_1:=x_1 - x_{1d}$.
\subsection{Adaptive speed observers}
\lab{subsec31}
%%%%%%%%%%%%%%%
%
In this subsection we present the two adaptive speed observers for the system \eqref{sys}.
\subsubsection{I$\&$I adaptive observer}
\lab{subsubsec311}
%%%%%%%5
The proof of the following proposition may be found in \cite[Proposition 1]{rom2025adaptive}.\footnote{In the algorithm given here we have added to extra tuning gains $\gamma_1,\gamma_2$, which gives us more flexibility to tune the observer. Their presence does not modify the validity of the convergence claim of \cite[Proposition 1]{rom2025adaptive}.}
\begin{prop}
\label{pro1}\em
Consider the system \eqref{sys}, assuming that $u$ ensures the states remain bounded. The I$\&$I adaptive velocity observer:
\begsubequ
\lab{iiobs1}
\begali{
\nonumber
\dot{x}_{2I}=&\frac{1}{J}u-\frac{ml_bg}{J}\sin(x_1)-\frac{1}{J}(\hat{\theta}_1+k_1)\hat{x}_2\\
&-\frac{1}{J}\hat{\theta}_2\tanh(\vartheta\hat{x}_2),\\
\dot{\theta}_{1I}=&\gamma_1  \frac{\vartheta}{k_1}\hat{x}_2\Big(\frac{k_1}{J}\hat{x}_2+\dot{x}_{2I}\Big),\\
\dot{\theta}_{2I}=&\gamma_2  \frac{\vartheta}{k_1}\tanh(\vartheta \hat{x}_2) \Big(\frac{k_1}{J}\hat{x}_2+\dot{x}_{2I}\Big),
}
\endsubequ
with the estimated velocity and parameters given by
\begsubequ
\lab{iiobs2}
\begali{
\label{eqn:adaptive2}
\hat{x}_2&=x_{2I}+\frac{k_1}{J}x_1,\\
\hat{\theta}_1&=\theta_{1I}-\gamma_1 \frac{\vartheta}{2k_1}\hat{x}_2^2,\\
\hat{\theta}_2&=\theta_{2I}-\gamma_2 \frac{1}{k_1}\ln\big(\cosh(\vartheta \hat{x}_2)\big),
}
\endsubequ
and tuning parameters $k_1,\gamma_1,\gamma_2 > 0$, guarantees that all signals remain bounded and
$$
\lim_{t\rightarrow \infty}[\hat{x}_2(t)-x_2(t)]=0,
$$
for all initial conditions $(x_1(0),x_2(0),x_{2I}(0)$,$\theta_{1I}(0)$,$\theta_{2I}(0))$\\$\in\mathbb{R}^5$.
\qed
\end{prop}

\subsubsection{SM adaptive observer}
\lab{subsubsec312}
%%%%%%%%%%5
We consider the SM adaptive observer reported in  \cite[Subsection 6.2]{SM1}, which is given by:
\begsubequ
\lab{smobs1}
\begali{
\dot{\hat{x}}_1(t)=&\hat{x}_2+\alpha_2\sqrt{|\tilde{x}_1|}\mathrm{sign}(\tilde{x}_1)\\
\nonumber
\dot{\hat{x}}_2(t)=&\frac{1}{J}u-\frac{ml_bg}{J}\sin(x_1)-\frac{\bar{\theta}_1}{J}\hat{x}_2\\
&-\frac{\bar{\theta}_2}{J}\tanh(\vartheta \hat{x}_2)+\alpha_1\mathrm{sign}(\tilde{x}_1)\\
\dot{\hat{\Delta}}_{\theta}=&\Gamma\boldsymbol\varphi[-\boldsymbol\varphi^\top \hat{\Delta}_\theta +\alpha_1\mathrm{sign}(\tilde{x}_1)]\\
\dot{\Gamma}=&-\Gamma\boldsymbol\varphi\boldsymbol\varphi^\top \Gamma,
}
\endsubequ
where $\bar {\boldsymbol \theta} \in \rea^2$ denotes an {\em a-priori fixed} estimate of the parameters $\boldsymbol \theta$, the constants $\alpha_1,\alpha_2$ are positive tuning gains, the regressor vector and the estimated parameters are defined as
$$
\boldsymbol\varphi:=-\begmat{ \hat{x}_2 \\ \tanh(\vartheta\hat{x}_2)},\;\hat{\boldsymbol\theta}:=J\bigg[\hat{\Delta}_{\theta}+\bar{\boldsymbol\theta}\bigg].
$$

It is claimed in \cite{SM1} that, if the signal $ \boldsymbol\varphi$ satisfies the excitation condition
\begequ
\lab{excsm}
\lim_{t\rightarrow \infty} \mathrm{inf} \frac{1}{t} \int_{0}^t  \boldsymbol\varphi(\sigma)\boldsymbol\varphi(\sigma)^\top d{\sigma}>0,
\endequ
and the controller gains $\alpha_1,\alpha_2$ are \textit{sufficiently large} then  the SM observer \eqref{smobs1} \textit{guarantees}
\begalis{
\lim_{t\rightarrow \infty}[\hat{x}_i(t)-x_i(t)]=0,\;i=1,2.\\
\lim_{t\rightarrow \infty}[\hat{\theta}_i(t)-\theta_i(t)]=0,\;i=1,2.
}
The SM design of \eqref{smobs1} belongs to the so-called third generation, according to the classification given in \cite{FRIetal}.

\subsection{Adaptive tracking controller}
\lab{subsec32}
%%%%%%%%%%%%%%%
%
We aim at achieving a closed-loop dynamics of the form
\begali{
\lab{con1}
\dot e_1 &= e_2 +\et,\\
\lab{con2}
\dot e_2 &= -k_p e_1 - k_v e_2 + \et,
}
where $e_1:=x_1-x_{1d},e_2:=x_2-\dot x_{1d}$, $k_p,k_v$ are positive constants and $\et \in \rea$ is a generic symbol for a signal {\em decaying to zero}. It is easy to see that, if $x_2$ is {\em measurable} and the friction parameters are {\em known}, the ideal control law
\begin{equation}\label{uid}
\begin{split}
u^\star=&\theta_1 x_2 + \theta_2 \tanh(\vartheta x_2)+ml_bg\sin(x_1)\\
&+J(\ddot x_{1d} -k_p e_1 - k_v(x_2 - \dot  x_{1d})),
\end{split}
\end{equation}
achieves this objective with $\et \equiv 0$. To achieve an implementable control, we propose, for both the I$\&$I and SM controllers, to use the estimated speed and parameters reported by the I$\&$I adaptive observer in Subsubsection \ref{subsubsec311}---in a certainty-equivalent manner---and those reported by the SM adaptive observer in Subsubsection \ref{subsubsec312}---again, in a certainty-equivalent manner---in \eqref{uid}, respectively. That is, we propose the adaptive control law:
\begin{equation}\label{u}
\begin{split}
u=&\hat \theta_1 \hat x_2 + \hat  \theta_2 \tanh(\vartheta \hat x_2)+ml_bg\sin(x_1)\\
&+J(\ddot  x_{1d} -k_p  e_1 -k_v (\hat x_2 - \dot  x_{1d})).
\end{split}
\end{equation}
It is easy to see that replacing the estimated quantities $\hat{(\cdot)}$ by $\tilde{(\cdot)}+(\cdot)$ we obtain $u=u^\star+\et,$ where
\begin{equation*}
\begin{split}
\et:=& \theta_1 \tilde x_2 + \tilde \theta_1(x_2+\tilde x_2)+ \tilde  \theta_2 \tanh(\vartheta (x_2+\tilde x_2))\\
&+ \theta_2 [\tanh(\vartheta (x_2+\tilde x_2)) - \tanh(\vartheta x_2)]-Jk_v \tilde x_2.
\end{split}
\end{equation*}
It is clear from the equation above that, if the estimated parameters $\hat \theta_i$ and estimated velocity $\hat x_2$ converge to their true values, we have that $\liminf \et(t)=0$, thus achieving the control objective, that is, ensuring the closed-loop structure \eqref{con1}–\eqref{con2}.\\

It is claimed in \cite{SM1} that, for the SM scheme discussed in Subsubsection \ref{subsubsec312}, the convergence required above follows under the standing excitation assumption \eqref{excsm}. On the other hand, it is shown in \cite[Proposition 2]{rom2025adaptive}, that to ensure parameter convergence in the I\&I scheme it is necessary to impose the following {\em excitation condition}.\footnote{As discussed in \cite{BARORT} the condition \eqref{newpe} is {\em strictly weaker} than the usual persistency of excitation assumption. It is also weaker than \eqref{excsm}.}\\

\noindent {\bf Assumption 1}
Consider the speed observer of Proposition \ref{pro1}, with the input signal given by \eqref{u}. Define the vector signal
\begequ
\lab{phi}
\boldsymbol\phi(t):=  \left[ \begin{array}{c} \hat x_2(t) \\ \tanh(\vartheta \hat x_2(t)) \end{array} \right].
\endequ
The reference signal $x_d(t)$ is such that the following condition holds true. There exist sequences of positive numbers $\{t_{k}\}$, $\{T_{k}\}$, and $\{\lambda_{k}\}$ such that $t_{k+1}\ge t_{k}+T_{k}$,  for $k=1,2,\dots$, $\inf\{T_{k}\}>0$, $\sup\{T_{k}\}<\infty$, and
\begequ
\lab{newpe}
\int_{t_{k}}^{t_{k}+T_{k}} \boldsymbol\phi(s)\boldsymbol\phi^{\top}(s) ds \ge \lambda_{k} I_2
\endequ
where $\sum_{k=1}^{\infty} \lambda_{k}^{2} = \infty.$
\qed \\

The proof of the proposition below may be found in  \cite[Proposition 2]{rom2025adaptive}.
\begin{prop}
\lab{pro2}\em
Consider the system \eqref{sys} in closed-loop with the control \eqref{u} where the estimated speed $\hat x_2$ and parameters $\hat \theta_i,\;i=1,2$, are generated by the I$\&$I adaptive speed observer of Proposition \ref{pro1}. If {\bf Assumption 1} holds true the {\em global tracking objective} is achieved. More precisely, we have that $\liminf \tilde \theta_i(t)=0,\;i=1,2$, consequently ensuring
$$
\liminf e_i(t)=0,\;i=1,2.
$$
\end{prop}
\begin{figure*}
  \centering
  \includegraphics[width=\textwidth]{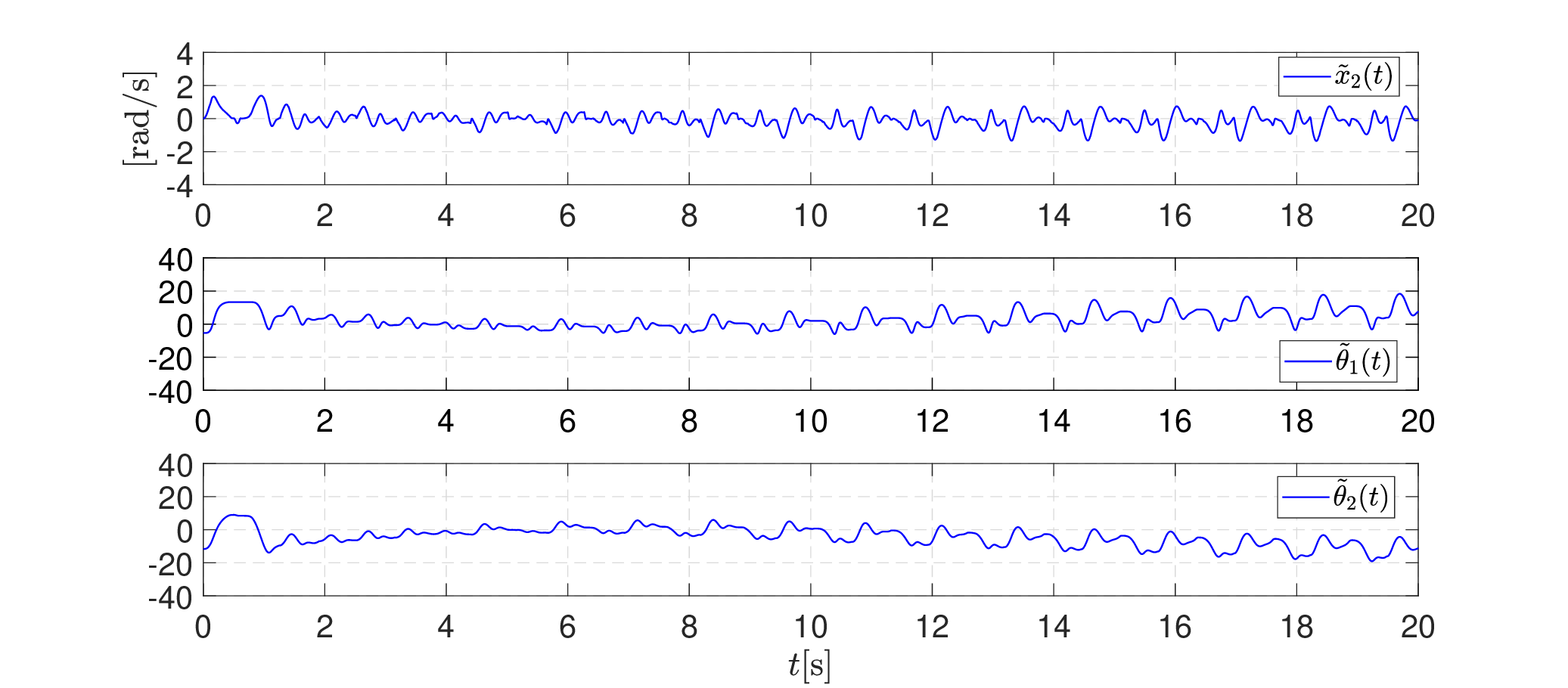}
  \caption{Behavior of the I$\&$I adaptive observer {\em error} signals in the open-loop test {\bf T1} for $u = 25 \sin(5t)$.}
    \label{fig3}
\end{figure*}
\begin{figure*}
  \centering
  \includegraphics[width=\textwidth]{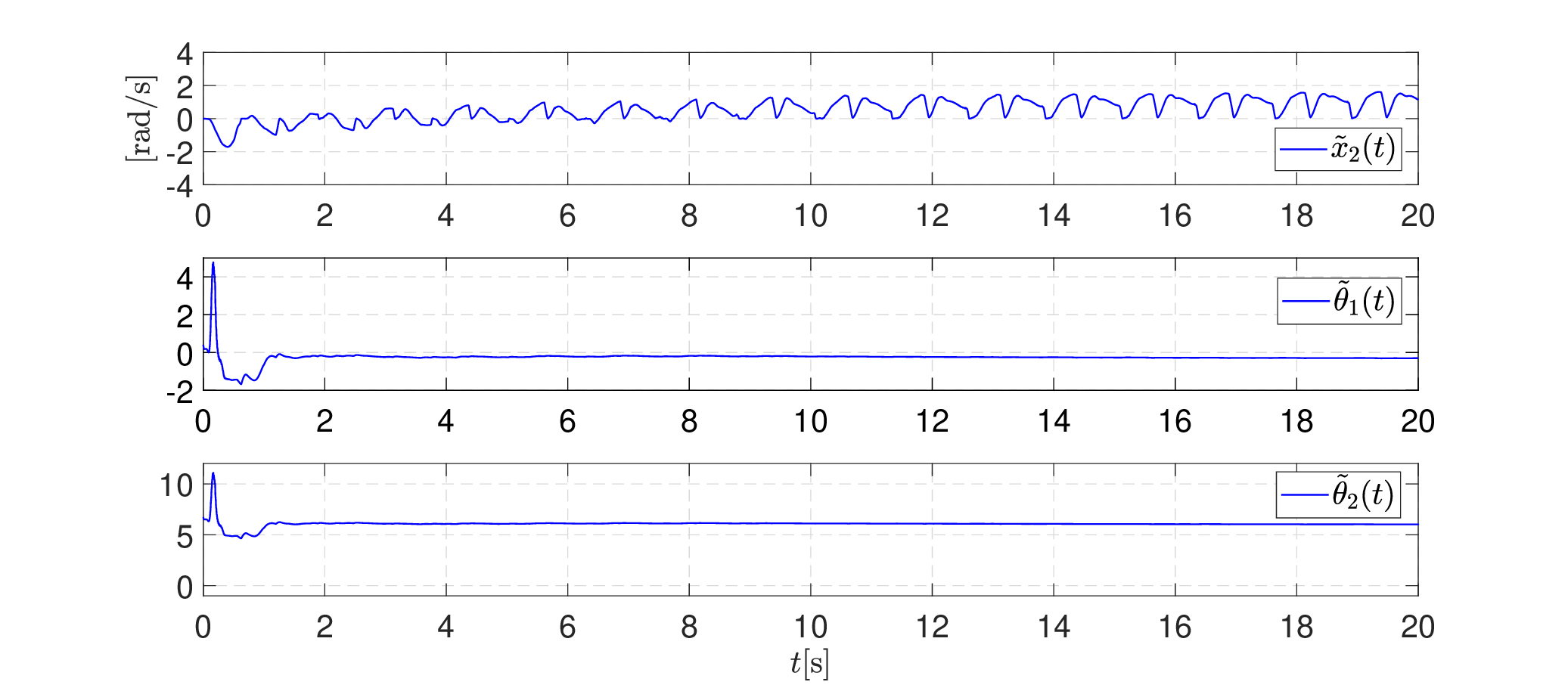}
  \caption{Behavior of the SM adaptive observer {\em error} signals in the open-loop test {\bf T1} for $u = 25 \sin(5t)$ with {\em excellent} prior knowledge of the parameters.}
    \label{fig4}
\end{figure*}
%%%%%%%%%%
 \section{Comparative Experiments: Open-loop estimation}
 \lab{sec4}
%%%%%%%%%%
%
In this section we give the experimental results for the I\&I and the SM adaptive observers described above when they are operating in {\em open-loop} for the two input signals given in test {\bf T1} of Subsection \ref{subsec22}.

For both input signals the initial conditions of the estimators were set to zero. That is, we set $\hat{x}_2(0)=0$, $\hat{\theta}_1(0)=0$, $\hat{\theta}_2(0)=0$ for the I\&I observer and for the SM observer we selected $\hat{x}_1(0) = 0$, $\hat{x}_2(0) = 0$, $\Gamma(0)=I_{2\times 2}$ and $\boldsymbol{\Delta}_\theta(0) = \begin{bmatrix}0 & 0\end{bmatrix}^\top$. Since the SM observer includes some a-priori estimates  of the parameters $\boldsymbol\theta$ we tried two different situations, which are described in the text below. Also, we tried different values for the various tuning parameters, explained also below.
\subsection{Results of test {\bf T1} for the input $u = 25 \sin(5t)$}
\lab{subsec41}
%%%%%%%%%%%%%
%
In this case the parameter of the $\tanh(\cdot)$ was set to $\vartheta=50$ for both observers. For the I\&I observer we set $k_1=1$, $\gamma_1=1,\gamma_2=1$. These gains were carefully selected according to the tuning guidelines provided in \cite{ReflectionsSM}. Since no tuning rules are available for the SM observer gains $\alpha_1$ and $\alpha_2$, we tried only small and large values.

Fig.~\ref{fig3} shows the behavior of the error signals for the I\&I observer. The figure shows that the estimate $\hat{x}_2(t)$ {\em converges} very rapidly to a small neighborhood of $x_2(t)$. Moreover, as expected from a ``non-exciting" input signal, the estimates $\hat{\theta}_1$ and $\hat{\theta}_2$ {\em do not converge} to their true values $\theta_1$ and $\theta_2$, however they remain sufficiently close to yield a good speed estimate.

\begin{figure*}
  \centering
  \includegraphics[width=0.8\textwidth]{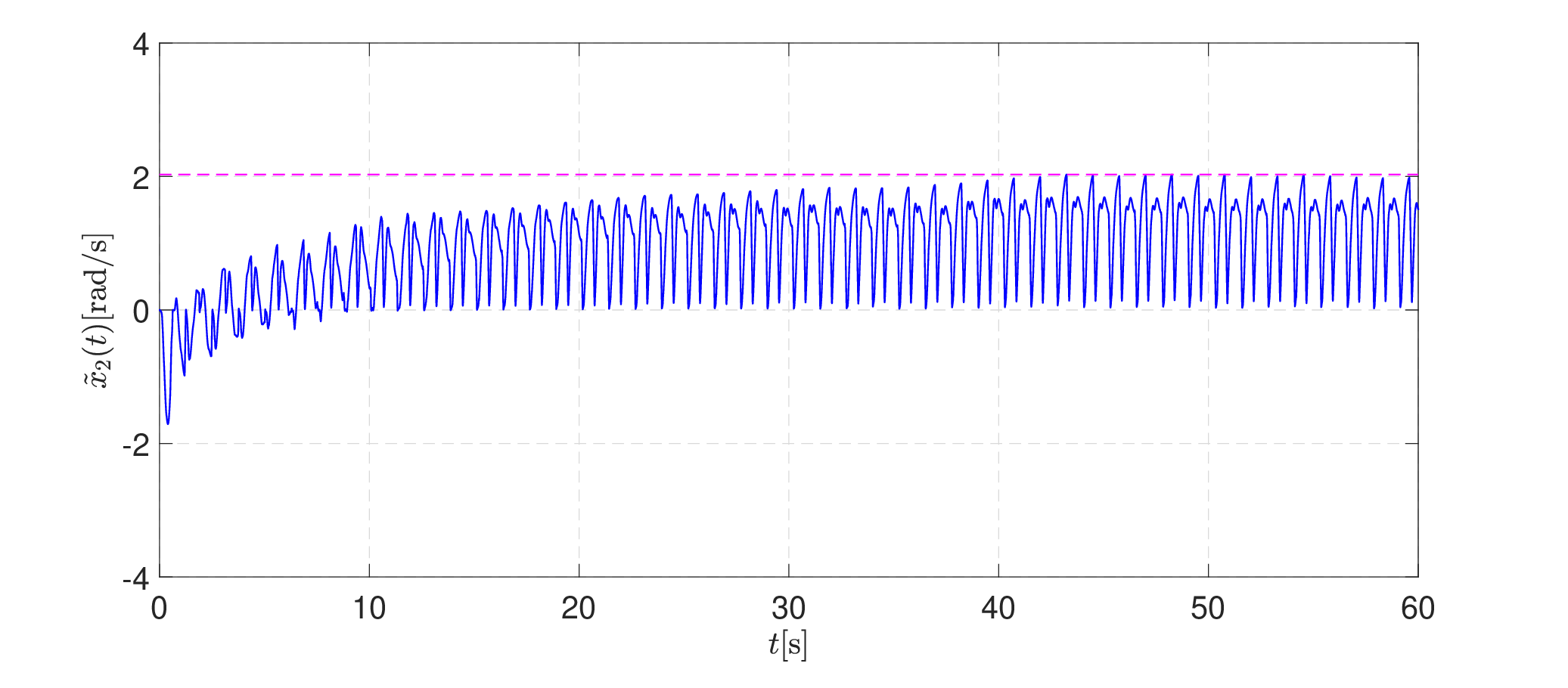}
  \caption{Speed {\em error} of SM observer in an extended simulation of Fig. \ref{fig4}.}
    \label{fig5}
\end{figure*}

For the SM observer we consider two different scenarios. First, the {\em ideal} situation where the parameters are {\em almost exactly} known, that is, we select  $\bar{\boldsymbol\theta} = \begin{bmatrix}7 & 15\end{bmatrix}^\top $, whereas the real values are ${\boldsymbol\theta}=\begin{bmatrix}7.5816 & 16.5981\end{bmatrix}$. The tuning gains were set to $\alpha_1 = 10$ and $\alpha_2 = 100$. The results are shown in Fig.~\ref{fig4}. As seen from the figure, even though the initial conditions of the estimates are very close to their true values, the parameter estimates {\em do not converge} and exhibit a large peak at the beginning. Moreover, the estimate $\hat{x}_2$ seems to grow {\em unbounded}. However, extending the length of the simulation as done in  Fig.~\ref{fig5}, we observe that it does not diverge, but exhibits an {\em inadmissible oscillatory behavior} in steady-state.

\begin{figure*}
  \centering
  \includegraphics[width=\textwidth]{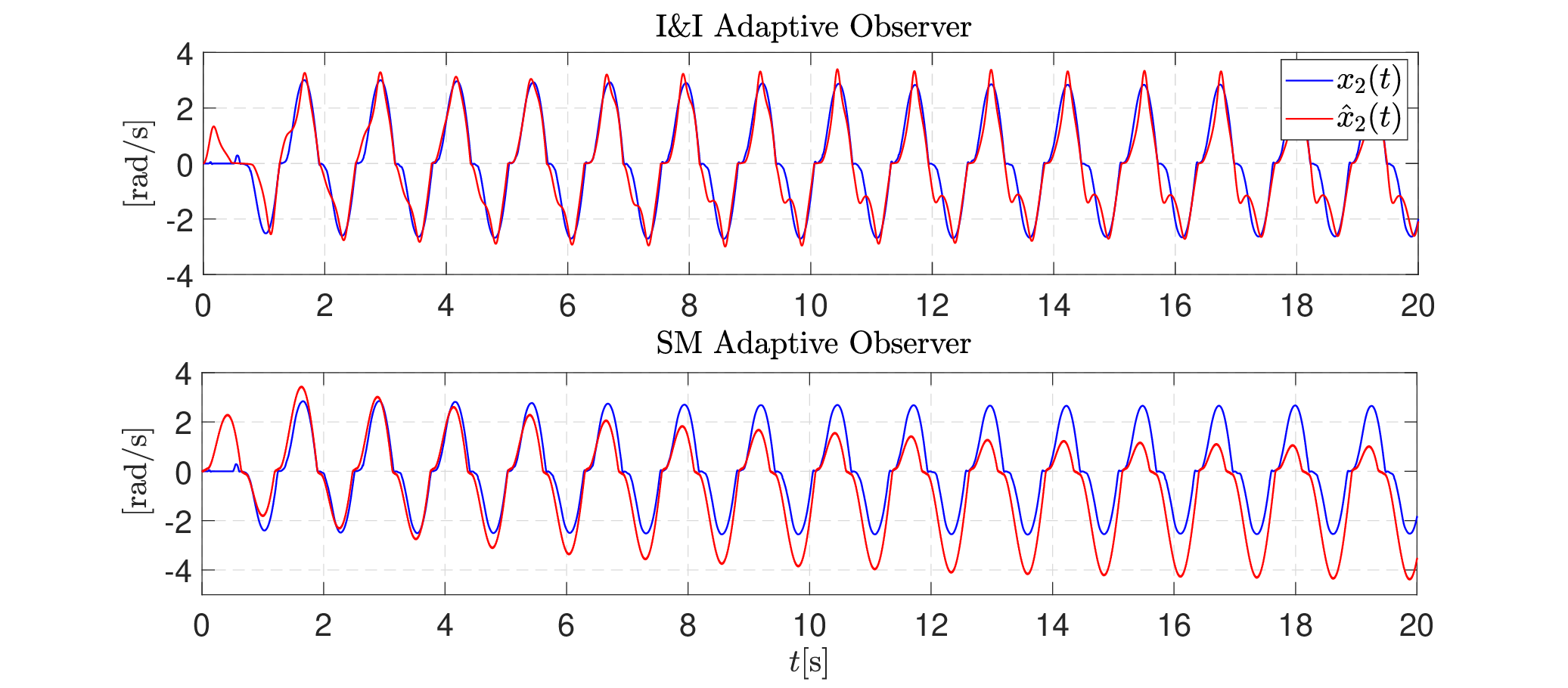}
\caption{Behavior of the {\em actual} and {\em estimated} velocity of the I$\&$I and the SM adaptive observer in the open-loop test {\bf T1} for $u = 25 \sin(5t)$.}
    \label{fig6}
\end{figure*}

The superiority of the I\&I observer over the SM one is better appreciated in Fig. \ref{fig6} where the {\em{actual}} speed signal is plotted together with its estimate for the I\&I and the SM observer, respectively. Notice the large steady-state deviation in the speed estimate of the SM observer.

%\begin{figure*}[h]
%  \centering
%  \includegraphics[width=\textwidth]{Figures/SM_O4_F2.eps}
%  \caption{Behavior  of the {\em actual} and {\em estimated} velocity of the SM adaptive observer in the open-loop test {\bf T1} for $u = 25 \sin(5t) \ \mathrm{[Nm]}$ with {\em excellent} prior knowledge of the parameters.}
%   \label{fig7}
%\end{figure*}

\begin{figure*}
  \centering
  \includegraphics[width=\textwidth]{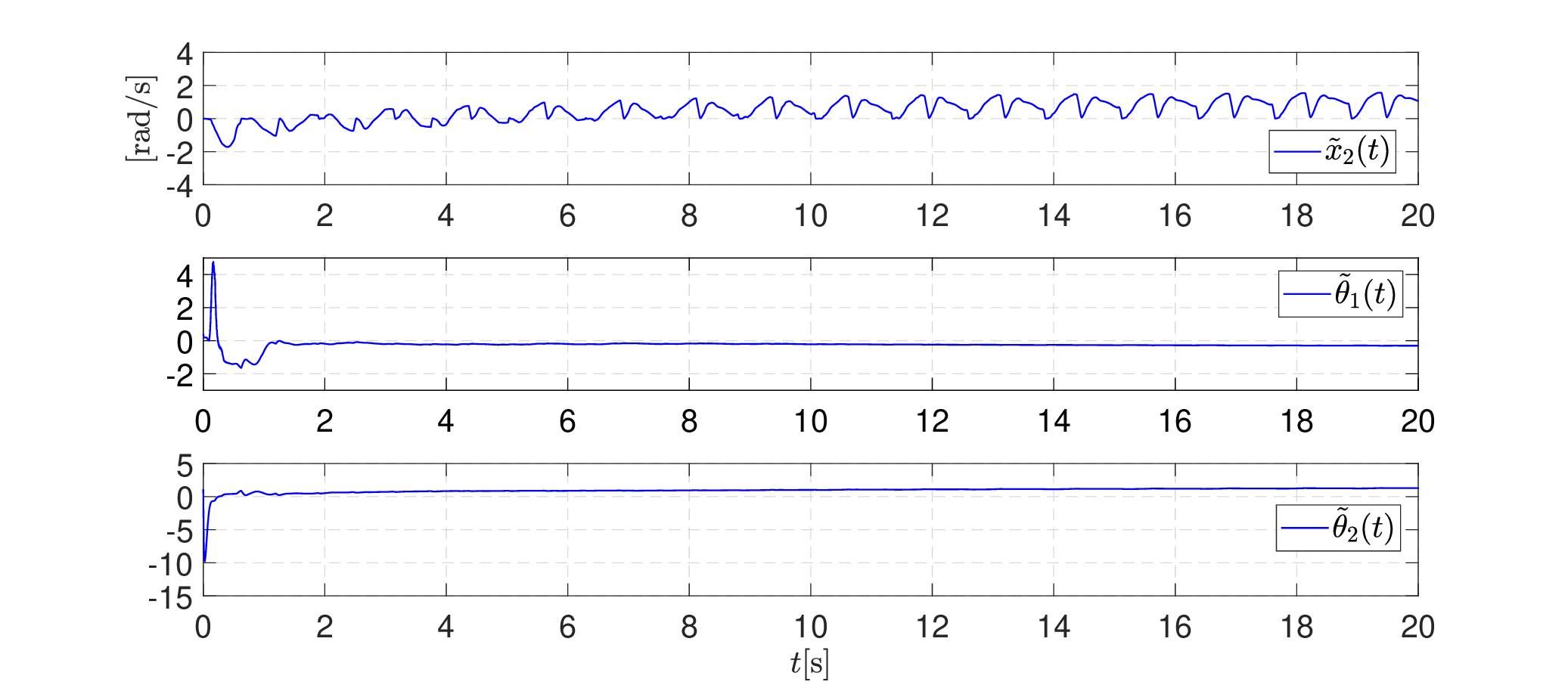}
  \caption{Behavior of the SM adaptive observer {\em error} signals in the open-loop test {\bf T1} for $u = 25 \sin(5t)$ {\em without} prior knowledge of the parameters.}
    \label{fig8}
\end{figure*}

We then considered the case of the SM observer when the nominal values $\bar{\boldsymbol{\theta}}$ are {\em far} from the real ones. Namely, we set $\bar{\boldsymbol\theta} = \begin{bmatrix}0.01 & 0.01\end{bmatrix}^\top $, keeping the same tuning gains $\alpha_1 = 10$ and $\alpha_2 = 100$. The results are shown in Fig. \ref{fig8}. Interestingly, even though the initial parameter estimate is far from their true values, it is seen that the estimates $\hat{\theta}_1$ and $\hat{\theta}_2$ behave relatively well, but still do not converge to their real values. Moreover, the initial peaks in the estimated parameters for the case of good and bad knowledge of the parameters are of similar height. This reveals that this term {\em plays no role} in the observer behavior. %Moreover, the estimate $\hat{x}_2$ exhibits {\em unbounded} behavior at the end of the experiment. \rom{Este si diverge?}.

\begin{figure*}
  \centering
  \includegraphics[width=\textwidth]{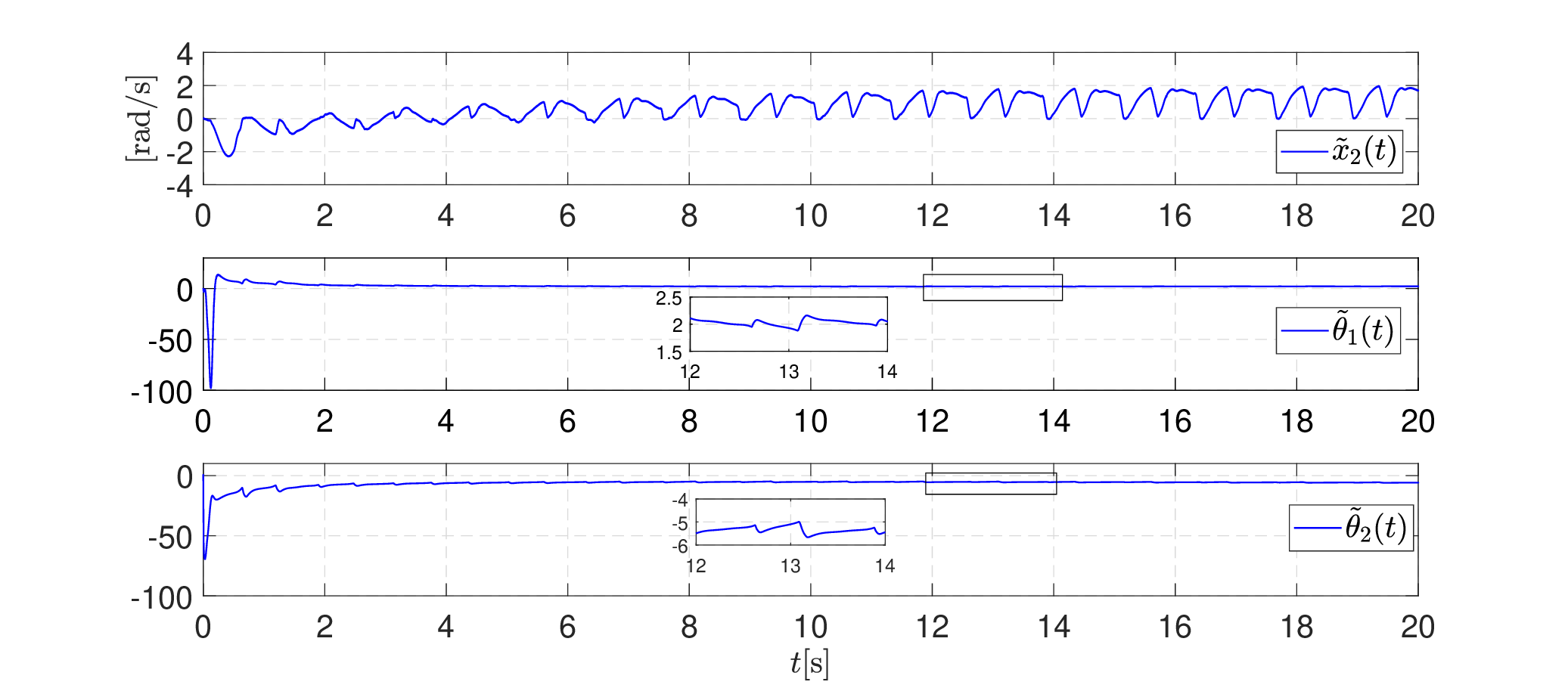}
  \caption{Behavior of the SM adaptive observer error signals in the open-loop test {\bf T1} for $u = 25 \sin(5t)$ with {\em high} gains $\alpha_1,\alpha_2$.}
    \label{fig9}
\end{figure*}

As a final test of the SM observer we tried it with {\em high tuning gains} $\alpha_1 = 100$ and $\alpha_2 = 1000$, and the {\em a priori} estimates {\em very close} to their true value, that is, $\bar{\boldsymbol\theta} = \begin{bmatrix}7 & 15\end{bmatrix}^\top $---with the experimental results given in Fig. \ref{fig9}. As seen in the figure, the estimated parameters approximate their true values more closely, showing better behavior than in the other scenarios, but still do not converge, even though their initial values were set closer to the true values. %Moreover, the estimate $\hat{x}_2$ still exhibits unbounded behavior. \rom{Este si diverge?}.

\subsection{Results of test {\bf T1} for the input $u = 14 \ \mathrm{sign} \big( \sin(\frac{\pi t}{3})\big)$}
\lab{subsec42}
%%%%%%%%%%%%%
%
In this case the parameter of the $\tanh(\cdot)$ was set to $\vartheta=500$ for both observers. For the I\&I observer we set $k_1=0.7$, $\gamma_1=0.7,\gamma_2=1$ and, similarly to the previous test, all initial conditions were set to zero. For the SM observer we set the {\em a priori} estimate of the parameters {\em almost equal} to their true value, namely, $\bar{\boldsymbol\theta} = \begin{bmatrix}7 & 15\end{bmatrix}^\top$, with the gains $\alpha_1 = 200$ and $\alpha_2 = 2000$ and initial conditions $\hat{x}_1(0) = 0$, $\hat{x}_2(0) = 0$, $\Gamma(0)=1000I_{2\times 2}$ and $\boldsymbol{\Delta}_\theta(0) = \begin{bmatrix}0 & 0\end{bmatrix}^\top $. %It is important to remark that for the SM observer it was necessary to put the initial parameter estimates very close to their true values and to increase the initial condition of the covariance matrix to obtain a stable response.  \rom{True?}

\begin{figure*}
  \centering
  \includegraphics[width=\textwidth]{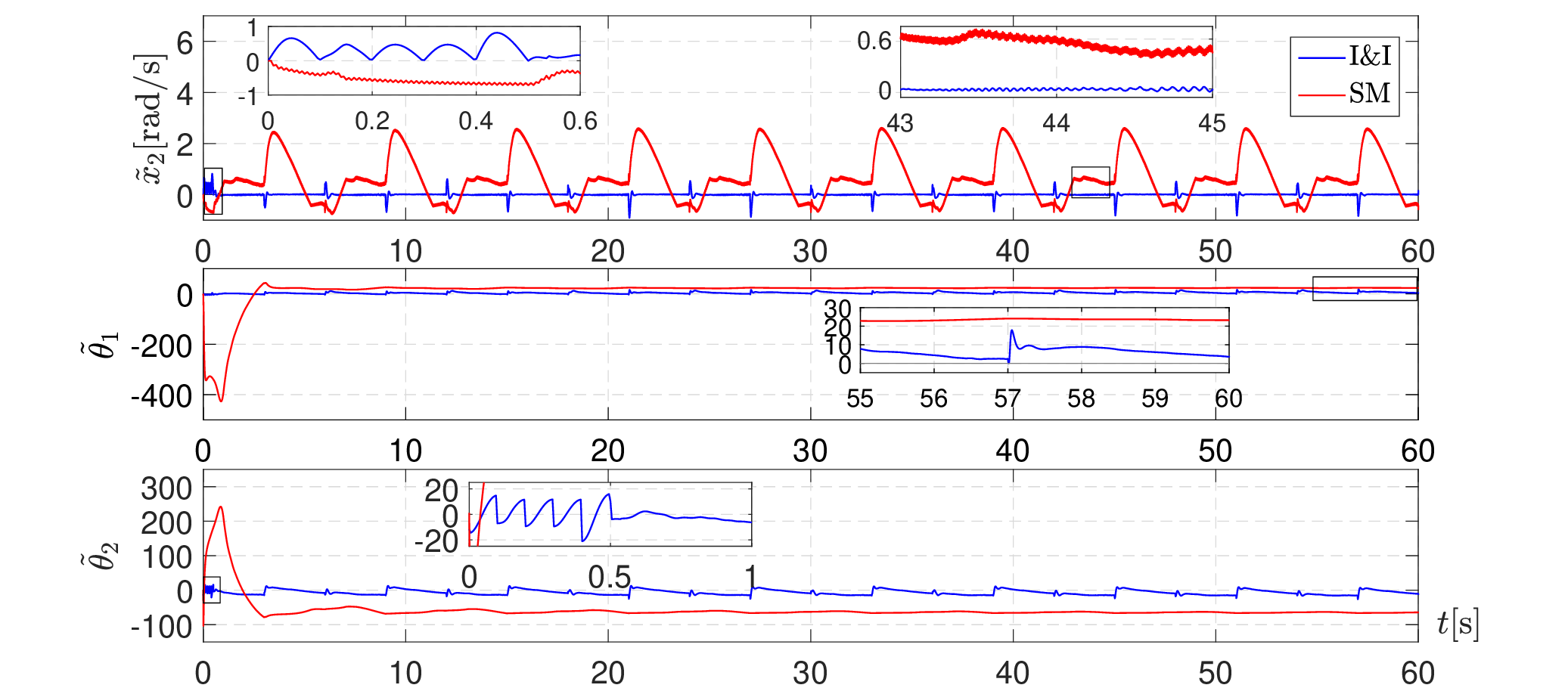}
  \caption{Behavior of {\em error} signals in the open-loop test {\bf T1} for $u=14 \ \mathrm{sign} \big( \sin(\frac{\pi t}{3})\big)$  for both observers, with {\em excellent} prior knowledge of the parameters in the SM one. }
    \label{fig10}
\end{figure*}

Fig.~\ref{fig10} shows the transient behavior of the error signals $\tilde{x}_2,\tilde{\theta}_1,\tilde{\theta}_2$ for both observers confirming the superior performance of the I$\&$I one. While the parameter estimate errors of the I\&I observer have small oscillations around zero, the ones of the SM observer exhibit a very large {\em steady-state bias}, which seems to be responsible for the poor behavior of the velocity observer. Moreover, as clearly depicted in the zoom of the time window $[43,45]$ the SM speed estimated has a strong {\em{chattering}}.

\begin{figure*}
  \centering
  \includegraphics[width=\textwidth]{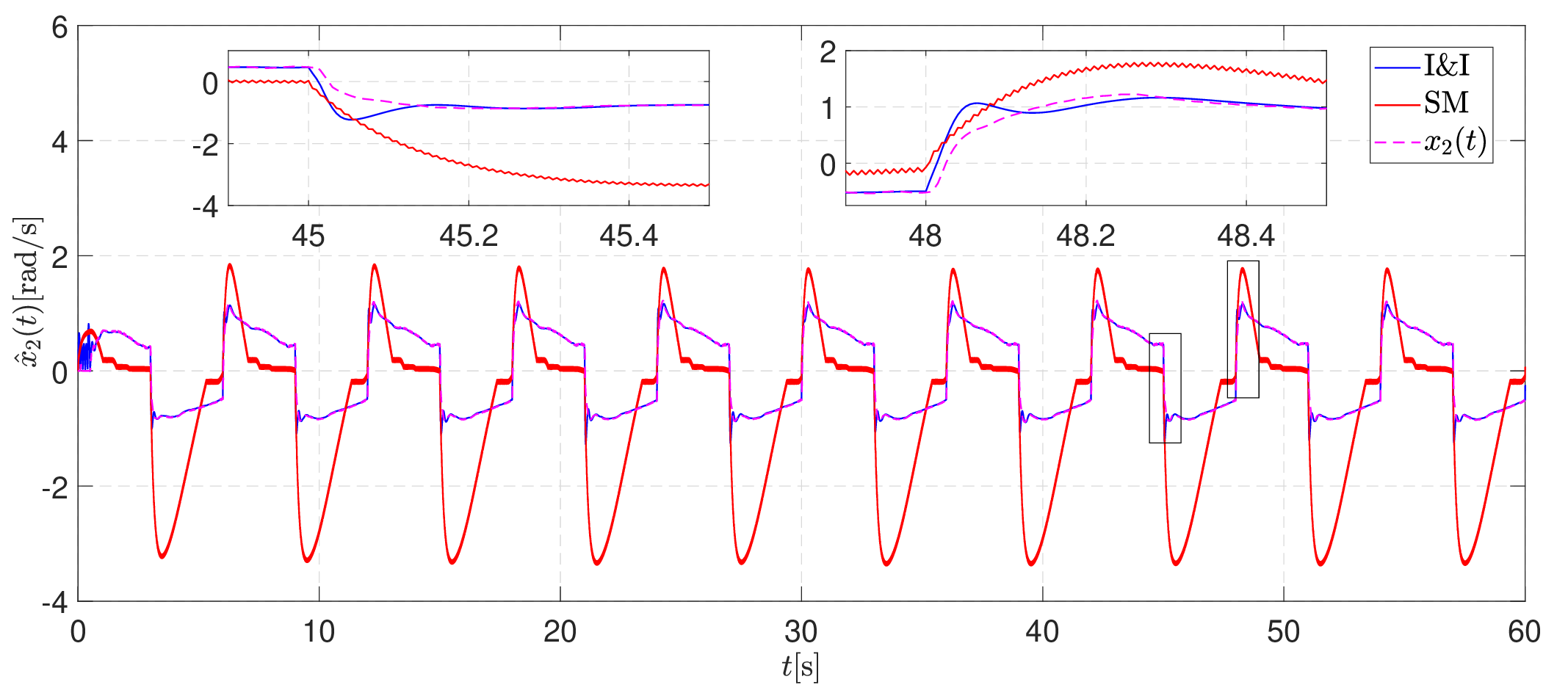}
  \caption{Behavior of the {\em actual} joint speed and its estimated values in the open-loop test {\bf T1} for $u=14 \ \mathrm{sign}\big( \sin(\frac{\pi t}{3})\big)$ for both observers, with {\em excellent} prior knowledge of the parameters in the SM one. }
    \label{fig11}
\end{figure*}

To further underscore this difference we show in Fig.~\ref{fig11} the {\em actual} link speed and its estimate for both observers. We bring to your attention the {\em boxed zooms}, where it is clearly seen that the I\&I observers closely tracks $x_2$, while the estimate of the SM observers is very far away. It is also noteworthy that the I$\&$I observer signals are {\em smooth}, in contrast to the SM observer states, particularly the estimate $\hat{x}_2(t)$, which exhibits a clear {\em chattering} effect.

%
%%%%%%%%%%%%%
\section{Comparative Experiments: Adaptive Tracking}
\lab{sec5}
%%%%%%%%%%%%5
%
The final test was designed to evaluate the performance of the tracking controller discussed in Subsection \ref{subsec32}, using the I\&I and the SM observers. It is proposed that the joint angle $x_1$ follows the desired trajectory:
\begin{equation}
\lab{xd}
x_d(t)=0.3\big[1-e^{-2.0t^3}\sin(7t) \big],
\end{equation}
which ensures that $\dot{x}_d(t)$ and $\ddot{x}_d(t)$ are bounded. The same experiment was also carried out for the {\em ``richer"} reference signal $x_d(t) = 0.5 \sin(5t)$, achieving {\em better} performance for both observers. In the interest of brevity, we omit the latter results, but they may be found in \cite{ReflectionsSMExperimental}.

 For these experiments we set the parameter $\vartheta = 330$ and fix the desired tracking error dynamics via $k_p = 1600$, $k_v = 1100$, which corresponds to the poles of the closed loop characteristic polynomial located at $p_1=-1.45$ and $p_2=-1098.54$. For the I$\&$I observer we used the gains $k_1 = 1$, $\gamma_1 = 0.03$, $\gamma_2 = 1$, and initial conditions $\hat{x}_2(0) = 0$, $\hat{\theta}_1(0) = 0$, $\hat{\theta}_2(0) = 0$. The controller gains were selected following the tuning guidelines in \cite{ReflectionsSM}. For the SM observer we used the almost exact parameters $\bar{\boldsymbol\theta} = \begin{bmatrix}7 & 15\end{bmatrix}^\top $ and initial conditions $\hat{x}_1(0) = 0$, $\hat{x}_2(0) = 0$, $\Gamma(0) = 100 I_2$, and $\boldsymbol{\Delta}\theta(0) = \begin{bmatrix}0 & 0\end{bmatrix}^\top $ and set the gains $\alpha_1 = 100$ and $\alpha_2 = 1500$.

\begin{figure*}[h]
  \centering
  \includegraphics[width=\textwidth]{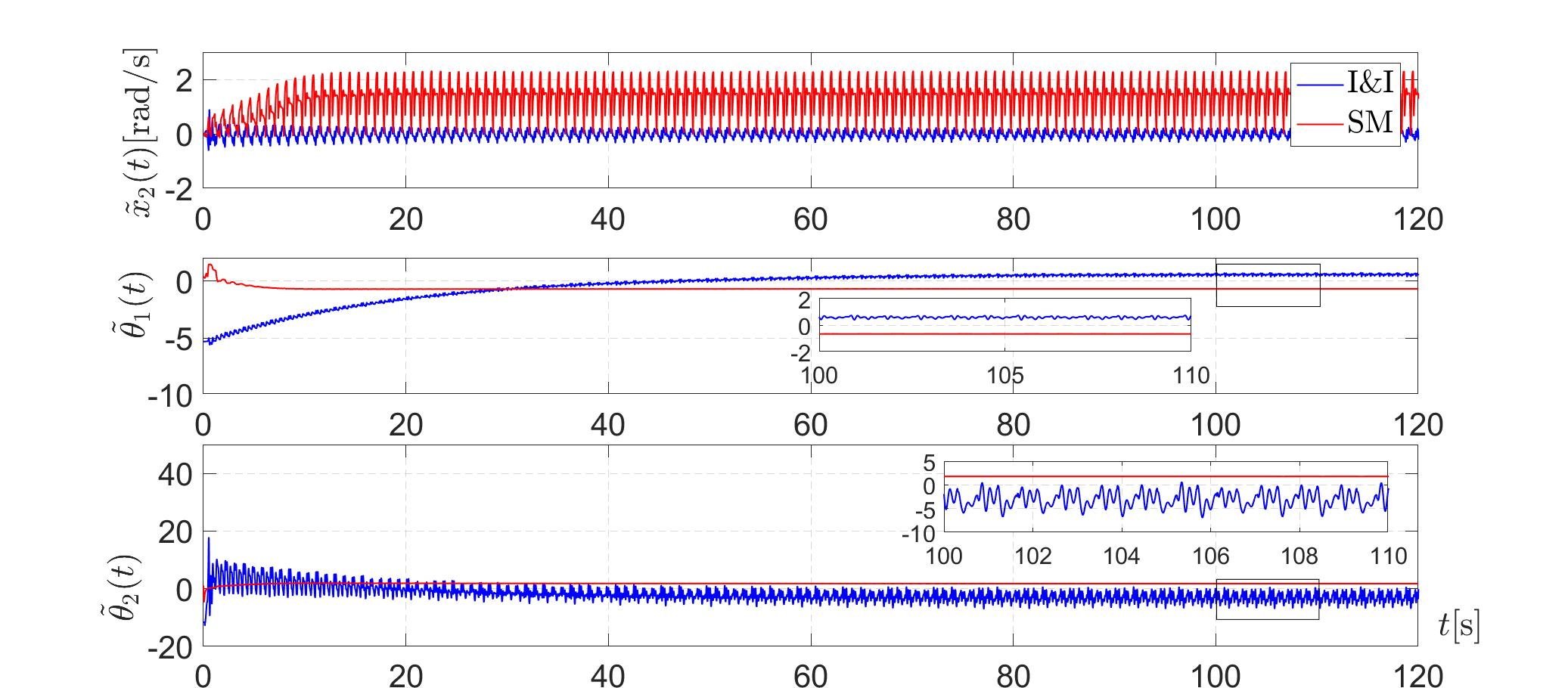}
  \caption{Behavior of the estimation {\em error} signals of the I$\&$I and the SM adaptive observers in the trajectory tracking experiment.}
    \label{fig12}
\end{figure*}

\begin{figure*}[h]
  \centering
  \includegraphics[width=\textwidth]{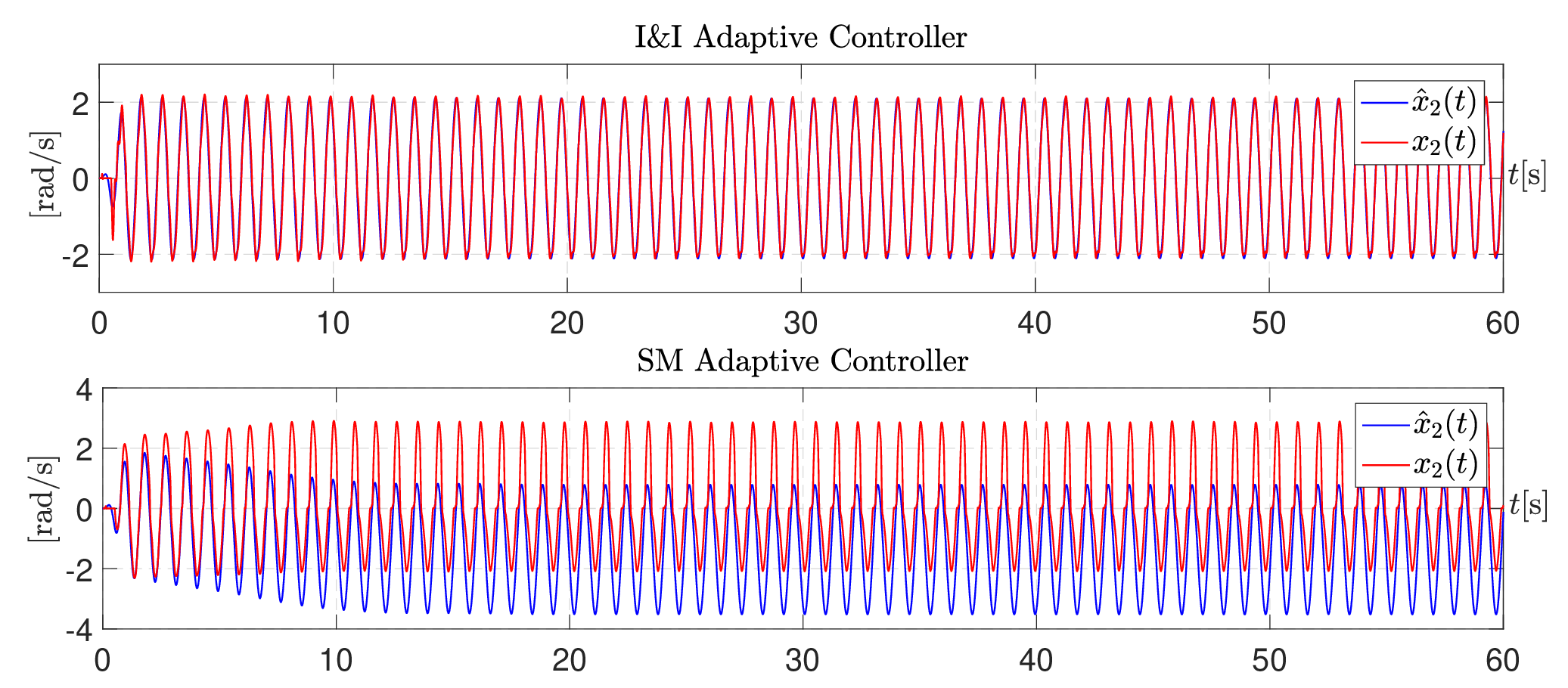}
  \caption{Behavior of the {\em actual} speed $x_2(t)$ and their estimate $\hat{x}_2(t)$ for the I$\&$I and the SM adaptive observers in the trajectory tracking experiment.}
    \label{fig13}
\end{figure*}

\begin{figure*}[h]
  \centering
  \includegraphics[width=\textwidth]{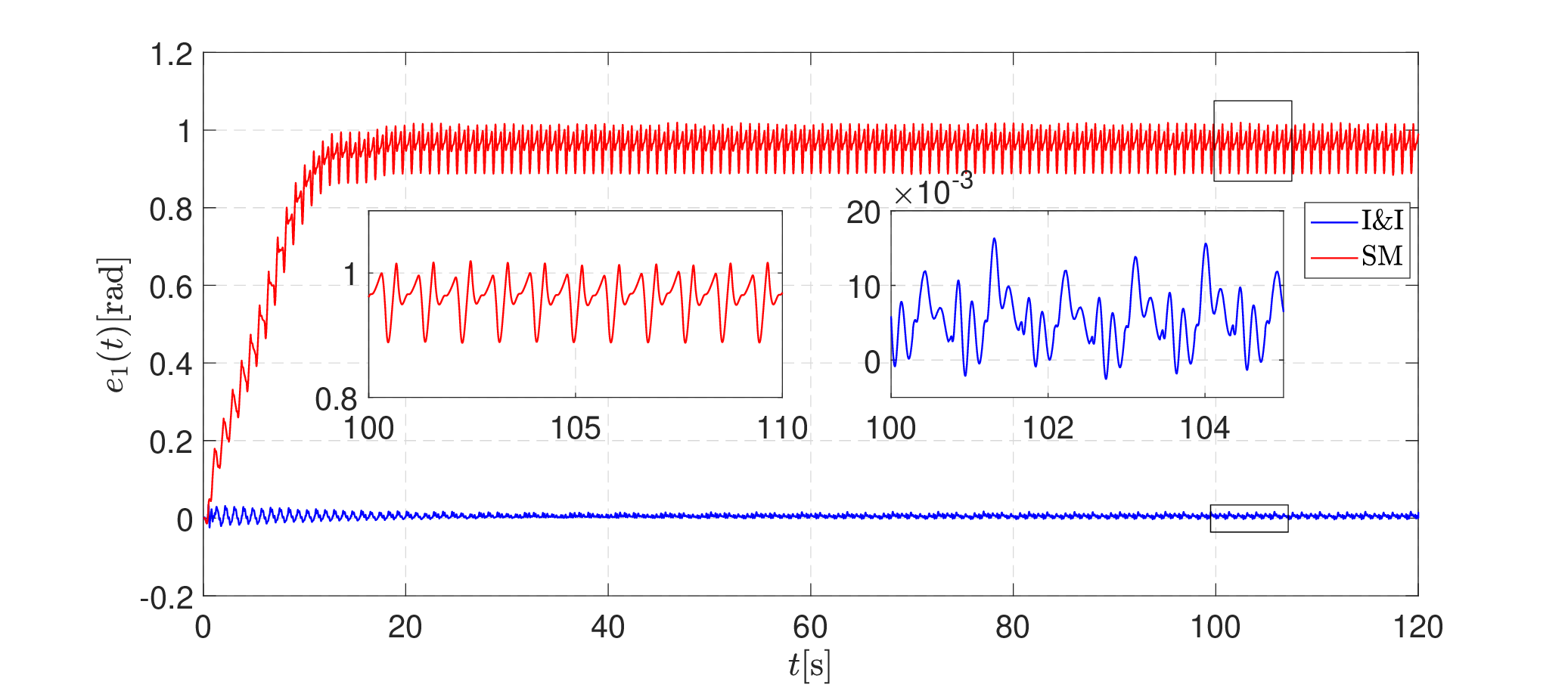}
  \caption{Behavior of the {\em tracking error} signal $e_1(t)$ of the I$\&$I and the SM adaptive controllers in the trajectory tracking experiment.}
    \label{fig14}
\end{figure*}

Fig.~\ref{fig12} shows the transient behavior of the {\em estimation error} signals, in Fig. \ref{fig13} we plot the {\em actual and estimated} speeds and in Fig. \ref{fig14} we show the {\em tracking error} signal for both observers. These figures clearly show the excellent behavior of the I\&I scheme. On the other hand, they reveal the  inadmissible performance of the SM one---unquestionably validating our claim of practical inadequacy of this technique.


\begin{figure*}[h]
  \centering
  \includegraphics[width=\textwidth]{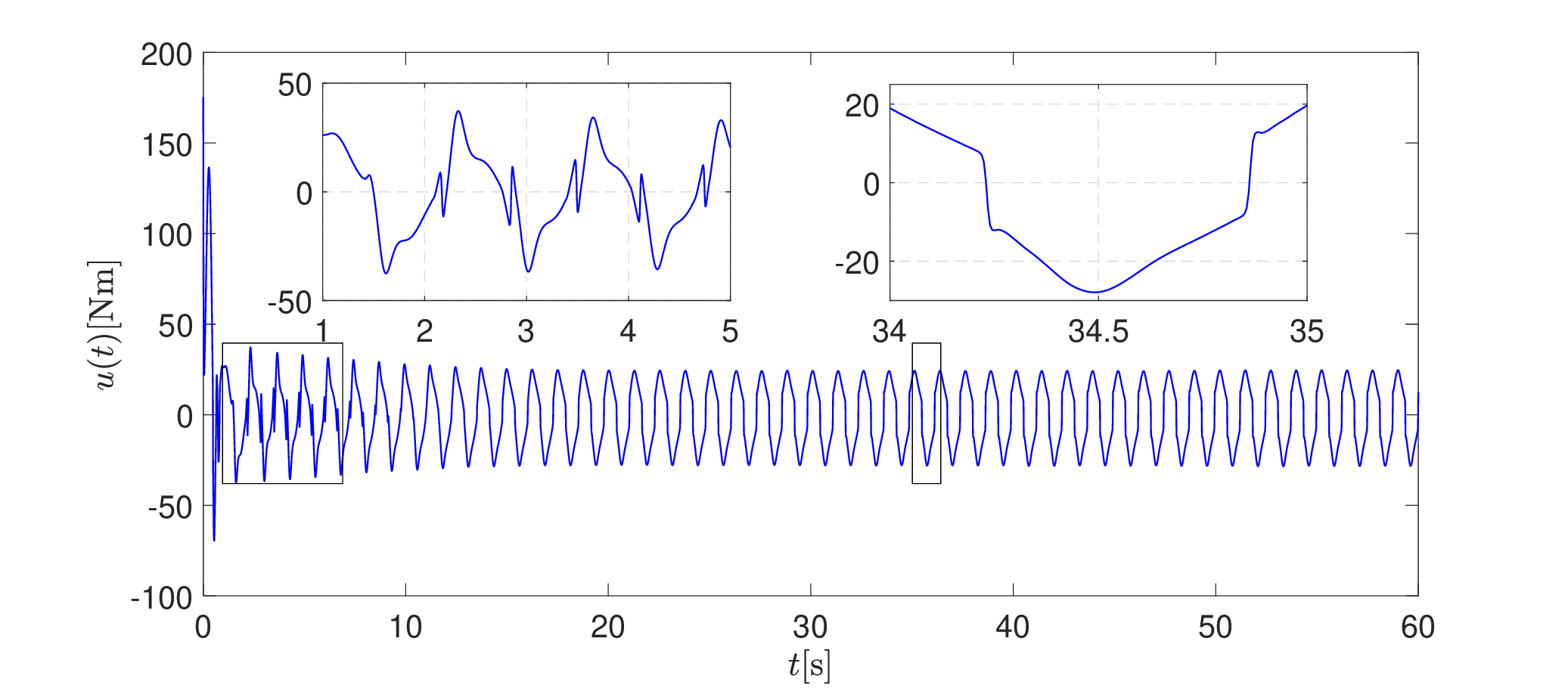}
  \caption{Behavior of the {\em control input} of the I$\&$I adaptive controller in the trajectory tracking experiment.}
    \label{fig15}
\end{figure*}

\begin{figure*}[h]
  \centering
  \includegraphics[width=\textwidth]{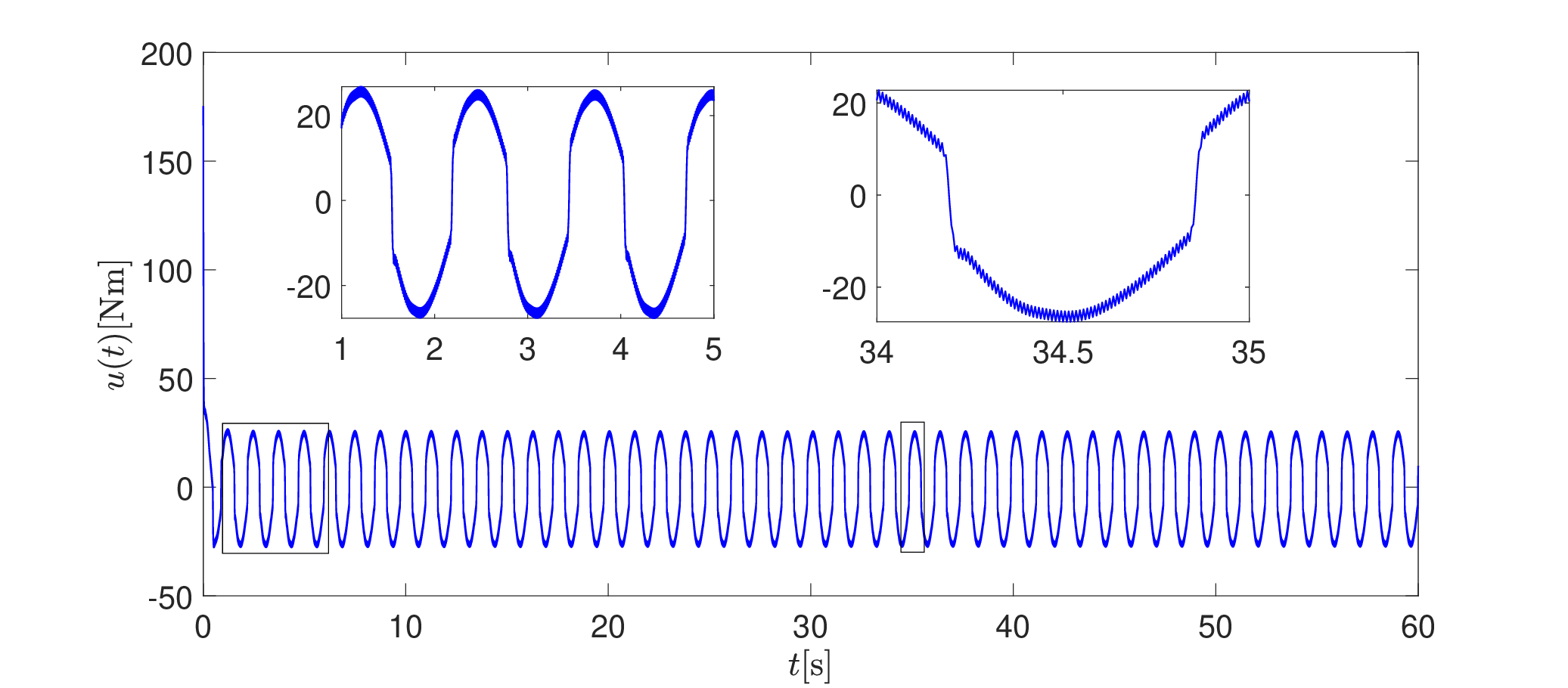}
  \caption{Behavior of the {\em control input} of the SM adaptive controller in the trajectory tracking experiment.}
    \label{fig16}
\end{figure*}

To further illustrate the deleterious effect of the HG injection implicit in SM designs, we plot the {\em control} signals for the I\&I and SM controllers in Figs. \ref{fig15} and \ref{fig16}, respectively. While the former exhibits a smooth behavior, the SM control plot shows a high-frequency component, inevitable in HG designs. It was observed that this  chattering phenomenon had the pernicious effect of heating the motor, that was sometimes stopped by the integrated temperature alarm system.

A {\em video} of the latter experimental test is available in the following link \url{https://youtu.be/bj3FBN8FLX4}. The desired reference is an oscillation of approximately $\pm 15^\circ$ around the downwards position of the pendulum. The video clearly shows that the I\&I controller closely, and smoothly, tracks this reference. On the other hand, the SM controller drifts approximately $30^\circ$ counterclockwise away from the hanging position. It should be noted also, that the high-frequency components of the control shown in Fig. \ref{fig16} induce a {\em motor noise} that is clearly audible in the video.

\section{The Narrative of Five Generations of SM Controllers}\label{sec6} %%%%%%%%%%%%5
In a recent paper \cite{FRIetal} the authors build up a narrative arguing that SM theory has evolved in five generations. The first one, pioneered by the work of Prof. Utkin in his ground-breaking 1981 monograph, where the controller is implemented with a {\em relay} which gives rise to significant chattering and the sliding surface design is restricted to have relative degree one. The second generation starts with the introduction of second order sliding modes, which in the twisting algorithm version boils down to using {\em two cascaded relay} functions. It is argued that the chattering can be ``reduced" adding an integrator in these schemes. The third generation is the super-twisting algorithm that now {\em splits the two relays} into one acting through the newly added integrator and the other one placed at the level of the control, which is ``scaled" with the fractional power of the absolute value of the state. Giving an interpretation of a state observer to this simple structure, it is argued that it can act as a ``robust exact SM {\em differentiator}". It is recognized that the chattering is not removed with this construction. It is argued that this scheme is insensitive to perturbations with {\em bounded derivative}---a rather restricted class of perturbations that rules out the typical step disturbance. The fourth generation arrives combining the ubiquitous relay controller with a nested structure of fractional powers of the absolute value of the sliding signal an its successive {\em derivatives}. The design requires the knowledge of an upperbound on a high order derivative of the sliding signal, which is not clear where this information will come from. The main objective of the fifth generation SM design is the addition of a {\em continuity feature}---arguing that due to the latter the chattering will be reduced, but not realizing that the performance degradation comes from the injection of HG through the always present relay operators.
% As indicated in Subsubsection \ref{subsubsec312} for the given application it assumes that, not only the system position and velocity, but even the acceleration and the control are {\em bounded}. It may be argued that the newer versions of SM would provide a better performance than the catastrophic one revealed above. To show that the problems highlighted in the experiment prevail in the new generations of SM designs, we present below a simulation study of a fifth generation design recently reported in the literature that clearly validates our argument.
%
\subsection{Example of a hydro-mechanical system}
\label{subsec61}
%%%%%%%%%%%%%%
 It may be argued that the newer versions of SM would provide a better performance than the SM design considered in the previous sections pertains to the third generation as per the classification of \cite{FRIetal}. To show that the problems highlighted in the experiment prevail in the new generations of SM designs, we present below a simulation study of a fifth generation design recently reported in the literature that clearly validates our argument.

We present in this subsection some simulations of a hydro-mechanical system \cite{hydro} with the {\em{novel}} SM adaptive observer proposed in \cite{SM2}. To highlight the performance degradation we compare it with the I$\&$I adaptive observer presented in  \cite{rom2025adaptive}. The objective of this study is to confirm that the essential practical limitations of SM designs remain in the newest versions of these schemes.

The dynamic model describing the linearized hydro-mechanical system has the form: \begin{equation}\label{eqn:hydro} \frac{d}{dt}\begin{bmatrix}x_1 \\ x_2 \\ x_3 \end{bmatrix}= \begin{bmatrix} x_2\\ -\theta_1x_2-\theta_2\tanh(\vartheta x_2)+a_1x_3\\ -a_2x_2-a_3x_3+u \end{bmatrix}, \end{equation} where $x_1$ and $x_2$ are the linear displacement and velocity of the cylinder, respectively; $x_3$ represents the differential load pressure. Following  \cite{hydro} it has been assumed that friction is present in the system modeled by stiction and Coulomb frictions, with $\theta_1$ and $\theta_2$ {\em{unknown}} parameters and $a_i,i=1,\ldots,3$, and $\vartheta$ {\em known} positive constants \footnote{Note that \eqref{eqn:hydro} coincides exactly with the linearized model of \cite{SM2} (equation (6)), being $a_1=\frac{A}{m}$, $a_2=\frac{4EA}{V_t}$, $a_3=\frac{4EC_{qp}}{V_t}$, $\frac{4EC_{q}}{V_t}=1$, $\delta_3(t)=0$, and $F_L(t)=0$, where $E$ is the bulk modulus, $V_t$ is the total hydraulic volume, $A$ is the average piston area, $\delta_3$ is a perturbation in the pressure load, and $F_L(t)$ represents an external load force. The constants $C_{qp}$ and $C_{q}$ are related to leakage in the cylinder. The function $\delta_2:=\sigma x_2-\theta_2\tanh(\vartheta x_2)$ is defined with $\theta_2>0$. Thus, it can be defined $\theta_1:=\sigma(\frac{1}{m}-1)$; provided that $m<1$, then $\theta_1>0$.}. In \cite{rom2025adaptive} an I$\&$I adaptive observer for the system \eqref{eqn:hydro} is presented, and its properties are given in the following.
\begin{prop}
\label{pro10}\em
Consider the system \eqref{eqn:hydro}, assuming that $u$ ensures the states remain bounded. The I$\&$I adaptive velocity observer: \begsubequ
\lab{iiobs1}
\begali{
\label{eqn:adaptive10}
\dot{x}_{2I}&=a_1\hat{x}_3-(\hat{\theta}_1+k_1)\hat{x}_2-\hat{\theta}_2\tanh(\vartheta\hat{x}_2),\\
\dot{\theta}_{1I}&=\frac{\vartheta}{k_1}\hat{x}_2\Big({k_1}\hat{x}_2+\dot{x}_{2I}\Big),\\
\dot{\theta}_{2I}&=\frac{\vartheta}{k_1}\tanh(\vartheta\hat{x}_2)\Big({k_1}\hat{x}_2+\dot{x}_{2I}\Big),
}
\endsubequ
with the estimated velocity $\hat{x}_2(t)$, the estimate $\hat{x}_3(t)$ of $x_3(t)$, and the estimated parameters given by
\begsubequ
\lab{iiobs2}
\begali{
\hat{x}_2&=x_{2I}+k_1x_1,\\
\hat{x}_3&=\int_{0}^{t}\big[-a_2\hat{x}_2(\tau)-a_3\hat{x}_3(\tau)+u(\tau)\big]d\tau,\\
\hat{\theta}_1&=\theta_{1I}- \frac{\vartheta}{2k_1}\hat{x}_2^2,\\
\label{eqn:adaptive20}
\hat{\theta}_2&=\theta_{2I}- \frac{1}{k_1}\ln\big(\cosh(\vartheta \hat{x}_2)\big),
}
\endsubequ
with tuning parameter $k_1>0$, guarantees that all signals remain bounded and
$$
\lim_{t\rightarrow \infty}\begin{bmatrix}\hat{x}_2(t)-x_2(t) \\ \hat{x}_3(t)-x_3(t) \end{bmatrix}=\bm{0}_2,
$$
for all initial conditions $(x_1(0),x_2(0),x_{2I}(0),x_3(0),\theta_{1I}(0),\theta_{2I}(0))\in\mathbb{R}^6$. \qed
\end{prop}

On the other hand, the SM Adaptive obsever of \cite{SM2} for the system \eqref{eqn:hydro} is given by:
\begsubequ
\lab{eqn:SMHydro1}
%\begin{align}
%\dot{\hat{\zeta}}_1=&-L^{\frac{1}{4}}c_1|\epsilon_1(t)|^{\frac{3}{4}}\mathrm{sign}(\epsilon_1(t))+\zeta_2,\\
%\dot{\hat{\zeta}}_2=&-L^{\frac{2}{4}}c_2|\epsilon_1(t)|^{\frac{2}{4}}\mathrm{sign}(\epsilon_1(t))+\zeta_3,\\
%\dot{\hat{\zeta}}_3=&-L^{\frac{3}{4}}c_3|\epsilon_1(t)|^{\frac{1}{4}}\mathrm{sign}(\epsilon_1(t))+\zeta_4+\frac{4EAC_q}{mV_t}u(t),\\ \dot{\hat{\zeta}}_4=&-Lc_4\mathrm{sign}(\epsilon_1(t))-\bigg( \frac{4\sigma EAC_q}{m^2V_t}+\frac{(4E)^2AC_qC_{qp}}{mV_t^2} \bigg)u(t),
%\end{align}
\begin{align}
\dot{\hat{\zeta}}_1=&-L^{\frac{1}{4}}c_1|\epsilon_1(t)|^{\frac{3}{4}}\mathrm{sign}(\epsilon_1(t))+\zeta_2,\\
\dot{\hat{\zeta}}_2=&-L^{\frac{2}{4}}c_2|\epsilon_1(t)|^{\frac{2}{4}}\mathrm{sign}(\epsilon_1(t))+\zeta_3,\\
\dot{\hat{\zeta}}_3=&-L^{\frac{3}{4}}c_3|\epsilon_1(t)|^{\frac{1}{4}}\mathrm{sign}(\epsilon_1(t))+\zeta_4+a_4u(t),\\ \dot{\hat{\zeta}}_4=&-Lc_4\mathrm{sign}(\epsilon_1(t))-a_5u(t),
\end{align}
\endsubequ
where $a_4$ and $a_5$ are some given constants stemming from the system model and $L$ and $c_i,i=1,\ldots,4$ are {\em{sufficiently}} large gains with $\epsilon_1:=\hat{\zeta}_1-x_1$.
 Several obscure signal boundedness assumptions on the system signals are then imposed, including the one that the system {\em acceleration}, as well as the {\em whole system state are bounded}. Note that the signal $\hat{\zeta}_2(t)$ represents an estimate of $x_2(t)$—that is, the velocity estimate—thus the velocity estimation error can be defined as $\tilde{\zeta}_2:=\hat{\zeta}_2-x_2(t)$. Also note that $\hat{\zeta}_3(t)$ is an estimate of the time derivative of $x_2(t)$, being $\zeta_3(t)=\dot{x}_2$, i.e., the system acceleration; hence, $\tilde{\zeta}_3=\hat{\zeta}_3(t)-\zeta_3(t)$ represents the acceleration estimation error. Moreover, in the SM adaptive observer, there is no estimate for the state $x_3(t)$, nor for the parameters $\theta_1$ and $\theta_2$. Although it is argued that this technique is ``model-based", this construction does not seem to reflect this claim. Finally, even more distressing condition, is imposed in Assumption 3, where boundedness of the derivative of the input of the tracking error dynamics is supposed. An observer-based controller is also proposed and it is claimed (in Theorem 1), that under the aforementioned boundedness assumptions and a {\em ``suitable" selection} of two tuning gains, there {\em exist sufficiently large} values for two more tuning gains that ensures the tracking error goes to zero in {\em ``some" finite time}. % %%%%%%%%%%%%%
\subsection{Simulation results}
\lab{subsec62}
In this subsection we present the results of numerical simulations of the hydro-mechanical system \eqref{eqn:hydro} using the I$\&$I adaptive observer \eqref{iiobs1} and \eqref{iiobs2} and compare them with the SM adaptive observer \eqref{eqn:SMHydro1}, using the following input signal:
\begin{equation}
u(t)=15 \sin(200t).
\end{equation}
The simulations were carried out in the Matlab/Simulink software using the ODE45 integration method, with an error tolerance of $1\times 10^{-6}$. The nominal dynamic parameters chosen for the hydro-mechanical system \eqref{eqn:hydro} were fixed as $a_1=1,a_2=1,a_3=1$, that is $m=0.5,\sigma=1,A=0.5,V_t=1,E=0.5,C_{q}=0.5,C_{qp}=0.5$ and $\theta_1=0.01,\theta_2=0.01$, and $\vartheta=100$, while the initial conditions were set as $[x_{1}(0) \ x_{2}(0) \ x_3(0) ]^T=[0\ 0 \ 0]^T$. For the I$\&$I adaptive observer, we set the initial observer conditions as $[x_1(0) \ x_2(0) \ x_{2I}(0) \ x_3(0) \ \theta_{1I}(0) \ \theta_{2I}(0)]^T=[0\ 0\ 0\ 0\ 0\ 0]^T$ and the gain $k_1=0.005$. For the SM adaptive observer, we set the initial conditions as $[\zeta_{1}(0) \ \zeta_{2}(0) \ \zeta_3(0) \ \zeta_4(0)]^T=[0\ 0 \ 0 \ 0]^T$.

\begin{figure*}
  \centering
  \includegraphics[width=\textwidth]{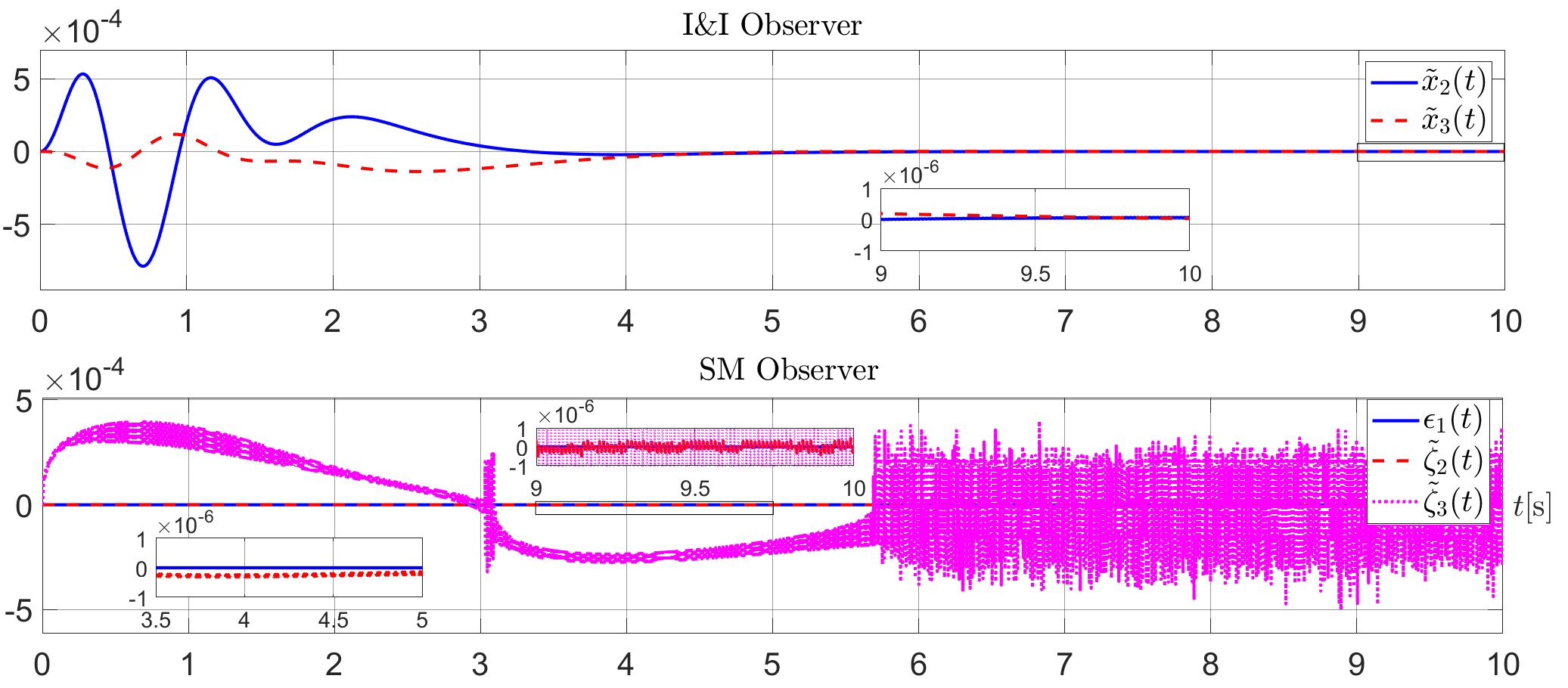} \caption{Behavior of the {\em error estimates} of the I$\&$I observer and the SM adaptive observer with the hydro-mechanical system \eqref{eqn:hydro} in the first simulation.} \label{fig20} \end{figure*}

The first simulation test was employed to tune the SM adaptive observer gains. Since, as usual, there is no clear guideline to tune the $L$ and $c_i$ observer gains, we selected as an initial point the gains reported in \cite{SM2}, that is, $L=650,c_1=3,c_2=4.16,c_3=3.06,c_4=1.1$. However, when running the simulation, {\em unbounded} trajectories were observed, indicating that the parameter $c_4$ was too high. Therefore, we decreased it significantly to $c_4=1.1 \times 10^{-4}$. The results are shown in Fig. \ref{fig20}, which shows that, the velocity estimation for the SM adaptive observer has a good performance; however, chattering appears in the signal $\tilde{\zeta}_3=\hat{\zeta}_3-\zeta_3$. As explained before, this signal can be seen as an approximation of $\dot{x}_2(t)$ and is expected to tend to zero in finite time.

\begin{figure*}
  \centering
  \includegraphics[width=\textwidth]{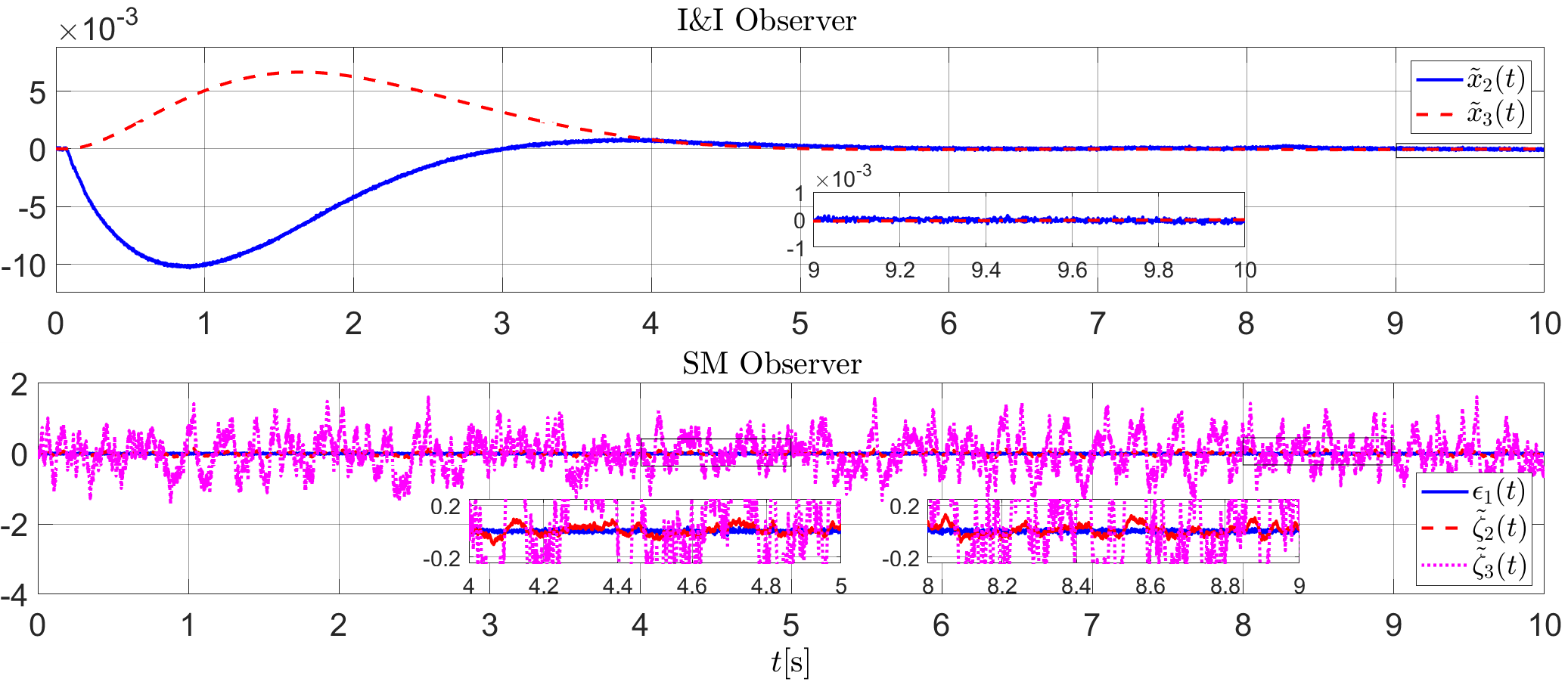}
  \caption{Behavior of the {\em error estimates} of the I$\&$I observer and the SM adaptive observer with the hydro-mechanical system \eqref{eqn:hydro} in the second simulation.} \label{fig21} \end{figure*}

Once both observers were tuned, we proceeded with the second simulation, whose purpose was to evaluate the performance of both adaptive observers—after gain tuning—under noise in the measurement of the plant states. Specifically, zero-mean Gaussian noise with a variance of $1\times10^{-4}$ was added to the input control channels of $x_1(t)$. The results are shown in Fig. \ref{fig21}. In this case, the SM adaptive observer exhibits a very strong chattering  in the error $\tilde{\zeta}_2$ and $\tilde{\zeta}_3$, while the I$\&$I adaptive observer has a smooth response. This confirms that the limitations of SM designs remain present, even in their most recent formulations.

 % % %%%%%%%%%%%%% \section{Concluding Remarks}

\section{Concluding Remarks}\label{sec7}
The experimental evidence presented in the paper unquestionably shows the practical implausability of high-gain injection-based techniques, for which the SM design reported in \cite{SM1} is a prototypical example. The task that was used to illustrate this fact is the implementation of an adaptive speed observer for the simplest mechanical system---a hanging pendulum---and the associated realization of an adaptive tracking controller. It is shown in \cite{rom2025adaptive} that this task has a simple solution invoking the well-known I\&I theory, a fact that was experimentally verified in this paper. The SM design that was implemented in the experimental facility belongs to the third generation and it may be argued that the newer versions of SM would provide a better performance. To show that the lack of robustness prevails in the new generations of SM designs, we present a simulation study of a fifth generation design recently reported in the literature that clearly validates our arguments. 
\par We wrap up this paper with a quote from Prof. Utkin's paper \cite{Utkin2016}:
\\
``\textit{At the moment the area is not specified where high order SM exhibits efficiency, since serious doubts and counterexamples were demonstrated}''.
\\

\section*{Acknowledgments}
This work was partially supported by SECIHTI under Grants CVU 1106239, by TecNM project numbers 22007.25-P, 22483.25-P and by Red Internacional de Control y Cómputo Aplicados del TecNM (RICCA / TecNM).

%\bmsection*{Financial disclosure}
%
%None reported.

\section*{Conflict of interest}

The authors declare no potential conflict of interests.
%\backmatter
%\bmsection*{Author contributions}
%
%This is an author contribution text. This is an author contribution text. This is an author contribution text. This is an author contribution text. This is an author contribution text.

%\bmsection*{Financial disclosure}
%
%None reported.

%\bibliographystyle{apalike}

%\bibliographystyle{apalike}
%\bibliography{biblio/Xbib}
\printbibliography
\end{document}